\pdfoutput=1

\documentclass[12pt,a4paper]{article}

\usepackage{ifthen} 
\newboolean{pdflatex}
\setboolean{pdflatex}{true} 

\newboolean{articletitles}
\setboolean{articletitles}{true} 

\newboolean{uprightparticles}
\setboolean{uprightparticles}{false} 

\newboolean{inbibliography}
\setboolean{inbibliography}{false} 

\usepackage{microtype}
\usepackage{lineno}  
\usepackage{xspace} 
\usepackage{caption} 

\usepackage{graphicx}  
\usepackage{color}
\usepackage{colortbl}
\graphicspath{{./figs/}} 

\usepackage{amsmath} 
\usepackage{amssymb}
\usepackage{amsfonts}
\usepackage{upgreek} 

\newcommand*\patchAmsMathEnvironmentForLineno[1]{%
\expandafter\let\csname old#1\expandafter\endcsname\csname #1\endcsname
\expandafter\let\csname oldend#1\expandafter\endcsname\csname
end#1\endcsname
 \renewenvironment{#1}%
   {\linenomath\csname old#1\endcsname}%
   {\csname oldend#1\endcsname\endlinenomath}%
}
\newcommand*\patchBothAmsMathEnvironmentsForLineno[1]{%
  \patchAmsMathEnvironmentForLineno{#1}%
  \patchAmsMathEnvironmentForLineno{#1*}%
}
\AtBeginDocument{%
\patchBothAmsMathEnvironmentsForLineno{equation}%
\patchBothAmsMathEnvironmentsForLineno{align}%
\patchBothAmsMathEnvironmentsForLineno{flalign}%
\patchBothAmsMathEnvironmentsForLineno{alignat}%
\patchBothAmsMathEnvironmentsForLineno{gather}%
\patchBothAmsMathEnvironmentsForLineno{multline}%
\patchBothAmsMathEnvironmentsForLineno{eqnarray}%
}

\usepackage{hyperref}    
\usepackage[all]{hypcap} 

\def\lhcb {\mbox{LHCb}\xspace}

\def\MagUp {\mbox{\em Mag\kern -0.05em Up}\xspace}

\ifthenelse{\boolean{uprightparticles}}%
{
 
 \def\Pgamma      {\ensuremath{\upgamma}\xspace}

 \def\Pmu         {\ensuremath{\upmu}\xspace}                 
 \def\Pnu         {\ensuremath{\upnu}\xspace}                 
                  
 \def\Ppi         {\ensuremath{\uppi}\xspace}                 
                  
 \def\Prho        {\ensuremath{\uprho}\xspace}

 \def\Ppsi        {\ensuremath{\uppsi}\xspace}

 \def\PDelta      {\ensuremath{\Delta}\xspace}                 
 \def\PXi      {\ensuremath{\Xi}\xspace}                 
 \def\PLambda      {\ensuremath{\Lambda}\xspace}                 
 \def\PSigma      {\ensuremath{\Sigma}\xspace}                 
 \def\POmega      {\ensuremath{\Omega}\xspace}                 
 \def\PUpsilon      {\ensuremath{\Upsilon}\xspace}                 
 
 \def\PB      {\ensuremath{\mathrm{B}}\xspace}                 
                  
 \def\PD      {\ensuremath{\mathrm{D}}\xspace}

 \def\PJ      {\ensuremath{\mathrm{J}}\xspace}                 
 \def\PK      {\ensuremath{\mathrm{K}}\xspace}

 \def\Pb      {\ensuremath{\mathrm{b}}\xspace}                 
 \def\Pc      {\ensuremath{\mathrm{c}}\xspace}                 
 \def\Pd      {\ensuremath{\mathrm{d}}\xspace}

 \def\Pi      {\ensuremath{\mathrm{i}}\xspace}

 \def\Ps      {\ensuremath{\mathrm{s}}\xspace}                 
 \def\Pt      {\ensuremath{\mathrm{t}}\xspace}                 
 \def\Pu      {\ensuremath{\mathrm{u}}\xspace}

}
{
 
 \def\Pgamma      {\ensuremath{\gamma}\xspace}

 \def\Pmu         {\ensuremath{\mu}\xspace}                 
 \def\Pnu         {\ensuremath{\nu}\xspace}                 
                  
 \def\Ppi         {\ensuremath{\pi}\xspace}                 
                  
 \def\Prho        {\ensuremath{\rho}\xspace}

 \def\Ppsi        {\ensuremath{\psi}\xspace}                 
                  
 \mathchardef\PDelta="7101
 \mathchardef\PXi="7104
 \mathchardef\PLambda="7103
 \mathchardef\PSigma="7106
 \mathchardef\POmega="710A
 \mathchardef\PUpsilon="7107
                  
 \def\PB      {\ensuremath{B}\xspace}                 
                  
 \def\PD      {\ensuremath{D}\xspace}

 \def\PJ      {\ensuremath{J}\xspace}                 
 \def\PK      {\ensuremath{K}\xspace}

 \def\Pb      {\ensuremath{b}\xspace}                 
 \def\Pc      {\ensuremath{c}\xspace}                 
 \def\Pd      {\ensuremath{d}\xspace}

 \def\Pi      {\ensuremath{i}\xspace}

 \def\Ps      {\ensuremath{s}\xspace}                 
 \def\Pt      {\ensuremath{t}\xspace}                 
 \def\Pu      {\ensuremath{u}\xspace}

}

\makeatletter
\ifcase \@ptsize \relax
  \newcommand{\miniscule}{\@setfontsize\miniscule{4}{5}}
\or
  \newcommand{\miniscule}{\@setfontsize\miniscule{5}{6}}
\or
  \newcommand{\miniscule}{\@setfontsize\miniscule{5}{6}}
\fi
\makeatother

\DeclareRobustCommand{\optbar}[1]{\shortstack{{\miniscule (\rule[.5ex]{1.25em}{.18mm})}
  \\ [-.7ex] $#1$}}

\def\mup        {{\ensuremath{\Pmu^+}}\xspace}
\def\mun        {{\ensuremath{\Pmu^-}}\xspace} 
\def\mumu       {{\ensuremath{\Pmu^+\Pmu^-}}\xspace}

\def\neu        {{\ensuremath{\Pnu}}\xspace}
\def\neub       {{\ensuremath{\overline{\Pnu}}}\xspace}
\def\neum       {{\ensuremath{\neu_\mu}}\xspace}
\def\neumb      {{\ensuremath{\neub_\mu}}\xspace}

\def\g      {{\ensuremath{\Pgamma}}\xspace}

\def\uquark    {{\ensuremath{\Pu}}\xspace}

\def\dquark    {{\ensuremath{\Pd}}\xspace}

\def\squark    {{\ensuremath{\Ps}}\xspace}

\def\cquark    {{\ensuremath{\Pc}}\xspace}

\def\bquark    {{\ensuremath{\Pb}}\xspace}

\def\tquark    {{\ensuremath{\Pt}}\xspace}

\def\pion   {{\ensuremath{\Ppi}}\xspace}
\def\piz    {{\ensuremath{\pion^0}}\xspace}

\def\pip    {{\ensuremath{\pion^+}}\xspace}
\def\pim    {{\ensuremath{\pion^-}}\xspace}
\def\pipm   {{\ensuremath{\pion^\pm}}\xspace}

\def\rhomeson {{\ensuremath{\Prho}}\xspace}
\def\rhoz     {{\ensuremath{\rhomeson^0}}\xspace}

\def\kaon    {{\ensuremath{\PK}}\xspace}
  \def\Kbar    {{\kern 0.2em\overline{\kern -0.2em \PK}{}}\xspace}

\def\KorKbar    {\kern 0.18em\optbar{\kern -0.18em K}{}\xspace}

\def\Kp      {{\ensuremath{\kaon^+}}\xspace}
\def\Km      {{\ensuremath{\kaon^-}}\xspace}

\def\KS      {{\ensuremath{\kaon^0_{\rm\scriptscriptstyle S}}}\xspace}

\def\Kstarz  {{\ensuremath{\kaon^{*0}}}\xspace}

  \def\Dbar    {{\kern 0.2em\overline{\kern -0.2em \PD}{}}\xspace}
\def\D       {{\ensuremath{\PD}}\xspace}

\def\DorDbar    {\kern 0.18em\optbar{\kern -0.18em D}{}\xspace}
\def\Dz      {{\ensuremath{\D^0}}\xspace}
\def\Dzb     {{\ensuremath{\Dbar{}^0}}\xspace}

\def\Dstarpm {{\ensuremath{\D^{*\pm}}}\xspace}

\def\B       {{\ensuremath{\PB}}\xspace}
\def\Bbar    {{\ensuremath{\kern 0.18em\overline{\kern -0.18em \PB}{}}}\xspace}

\def\BorBbar    {\kern 0.18em\optbar{\kern -0.18em B}{}\xspace}

\def\Bu      {{\ensuremath{\B^+}}\xspace}
\def\Bub     {{\ensuremath{\B^-}}\xspace}
\def\Bp      {{\ensuremath{\Bu}}\xspace}
\def\Bm      {{\ensuremath{\Bub}}\xspace}
\def\Bpm     {{\ensuremath{\B^\pm}}\xspace}

\def\Bd      {{\ensuremath{\B^0}}\xspace}
\def\Bs      {{\ensuremath{\B^0_\squark}}\xspace}

\def\jpsi     {{\ensuremath{{\PJ\mskip -3mu/\mskip -2mu\Ppsi\mskip 2mu}}}\xspace}

  \def\Y#1S{\ensuremath{\PUpsilon{(#1S)}}\xspace}

\def\Lbar        {{\ensuremath{\kern 0.1em\overline{\kern -0.1em\PLambda}}}\xspace}
\def\LorLbar    {\kern 0.18em\optbar{\kern -0.18em \PLambda}{}\xspace}

\def\BF         {{\ensuremath{\cal B}}\xspace}

\def\BR         {\BF}
\newcommand{\decay}[2]{\ensuremath{#1\!\to #2}\xspace}         

\def\to                 {\ensuremath{\rightarrow}\xspace}

\def\qsq       {{\ensuremath{q^2}}\xspace}

\def\CP                {{\ensuremath{C\!P}}\xspace}

\def\Vtd  {{\ensuremath{V_{\tquark\dquark}}}\xspace}

\def\Vts  {{\ensuremath{V_{\tquark\squark}}}\xspace}
\def\Vub  {{\ensuremath{V_{\uquark\bquark}}}\xspace}

\def\Vtb  {{\ensuremath{V_{\tquark\bquark}}}\xspace}
\def\Vuds  {{\ensuremath{V_{\uquark\dquark}^\ast}}\xspace}

\def\Vtds  {{\ensuremath{V_{\tquark\dquark}^\ast}}\xspace}

\def\AT#1     {\ensuremath{A_{\mathrm{T}}^{#1}}\xspace}           

\def\C#1      {\ensuremath{\mathcal{C}_{#1}}\xspace}                       
\def\Cp#1     {\ensuremath{\mathcal{C}_{#1}^{'}}\xspace}                    
\def\Ceff#1   {\ensuremath{\mathcal{C}_{#1}^{\mathrm{(eff)}}}\xspace}        
\def\Cpeff#1  {\ensuremath{\mathcal{C}_{#1}^{'\mathrm{(eff)}}}\xspace}       
\def\Ope#1    {\ensuremath{\mathcal{O}_{#1}}\xspace}                       
\def\Opep#1   {\ensuremath{\mathcal{O}_{#1}^{'}}\xspace}                    

\newcommand{\tev}{\ifthenelse{\boolean{inbibliography}}{\ensuremath{~T\kern -0.05em eV}\xspace}{\ensuremath{\mathrm{\,Te\kern -0.1em V}}}\xspace}
\newcommand{\gev}{\ensuremath{\mathrm{\,Ge\kern -0.1em V}}\xspace}
\newcommand{\mev}{\ensuremath{\mathrm{\,Me\kern -0.1em V}}\xspace}
\newcommand{\kev}{\ensuremath{\mathrm{\,ke\kern -0.1em V}}\xspace}
\newcommand{\ev}{\ensuremath{\mathrm{\,e\kern -0.1em V}}\xspace}
\newcommand{\gevc}{\ensuremath{{\mathrm{\,Ge\kern -0.1em V\!/}c}}\xspace}
\newcommand{\mevc}{\ensuremath{{\mathrm{\,Me\kern -0.1em V\!/}c}}\xspace}
\newcommand{\gevcc}{\ensuremath{{\mathrm{\,Ge\kern -0.1em V\!/}c^2}}\xspace}
\newcommand{\gevgevcccc}{\ensuremath{{\mathrm{\,Ge\kern -0.1em V^2\!/}c^4}}\xspace}
\newcommand{\mevcc}{\ensuremath{{\mathrm{\,Me\kern -0.1em V\!/}c^2}}\xspace}

\def\mum  {\ensuremath{{\,\upmu\rm m}}\xspace}

\def\invfb   {\ensuremath{\mbox{\,fb}^{-1}}\xspace}

\newcommand{\stat}{\ensuremath{\mathrm{\,(stat)}}\xspace}
\newcommand{\syst}{\ensuremath{\mathrm{\,(syst)}}\xspace}

\def\deriv {\ensuremath{\mathrm{d}}}

\def\gsim{{~\raise.15em\hbox{$>$}\kern-.85em
          \lower.35em\hbox{$\sim$}~}\xspace}
\def\lsim{{~\raise.15em\hbox{$<$}\kern-.85em
          \lower.35em\hbox{$\sim$}~}\xspace}

\def\ptot       {\mbox{$p$}\xspace}
\def\pt         {\mbox{$p_{\rm T}$}\xspace}

\def\evtgen     {\mbox{\textsc{EvtGen}}\xspace}

\def\geant      {\mbox{\textsc{Geant4}}\xspace}

\def\photos     {\mbox{\textsc{Photos}}\xspace}

\def\pythia     {\mbox{\textsc{Pythia}}\xspace}

\def\tell1  {TELL1\xspace}
\def\ukl1   {UKL1\xspace}

\newcommand{\ie}{\mbox{\itshape i.e.}\xspace}

\usepackage{cite} 
\usepackage{mciteplus}

\def\dqsq {\ensuremath{\deriv\qsq}\xspace}

\def\Burhomumu {\decay{\Bu}{\rho^+(\pip\piz)\mumu}}
\def\Bdrhomumu {\decay{\Bd}{\rho^0(\pip\pim)\mumu}}
\def\Bsfzmumu {\decay{\Bs}{f_0(\pip\pim)\mumu}}

\def\BuDzpimumu {\decay{\Bu}{\Dzb(\pip\mun\neumb)\mup\neum}}

\def\Bupimm {\decay{\Bu}{\pip\mumu}}

\def\BuJpsipi {\decay{\Bu}{\jpsi(\mumu)\pip}}
\def\BuKmm {\decay{\Bu}{\Kp\mumu}}
\def\BuJpsiK {\decay{\Bu}{\jpsi(\mumu)\Kp}}

\def\Bpmpimm {\decay{\Bpm}{\pipm\mumu}}

\def\pimumu {\pip\mumu}

\def\kmumu {\Kp\mumu}

\def \btodll {\ensuremath{\bquark\to\dquark \ell^+ \ell^-}\xspace}

\def \Bptopimumu {\ensuremath{\Bp\to\pip\mumu}\xspace}
\def \Bmtopimumu {\ensuremath{\Bm\to\pim\mumu}\xspace}
\def \Bppimm {\ensuremath{\Bp\to\pip\mumu}\xspace}
\def \Bmpimm {\ensuremath{\Bm\to\pim\mumu}\xspace}

\def \acp {\ensuremath{\mathcal{A}_{\CP}}\xspace}
\def \ap {\ensuremath{\mathcal{A}_\mathrm{P}}\xspace}
\def \araw {\ensuremath{\mathcal{A}_\mathrm{RAW}}\xspace}
\def \adet {\ensuremath{\mathcal{A}_\mathrm{DET}}\xspace}

\def \refprevsec #1 {(Section #1,\xspace\cite{Ciezarek:1393751})\xspace}
\def \refprevfig #1 {(Figure #1,\xspace\cite{Ciezarek:1393751})\xspace}
\def \refprevtab #1 {(Table #1,\xspace\cite{Ciezarek:1393751})\xspace}

\usepackage{longtable} 
\usepackage{float}

\begin{document}

\renewcommand{\thefootnote}{\fnsymbol{footnote}}
\setcounter{footnote}{1}


\begin{titlepage}
\pagenumbering{roman}

\vspace*{-1.5cm}
\centerline{\large EUROPEAN ORGANIZATION FOR NUCLEAR RESEARCH (CERN)}
\vspace*{1.5cm}
\noindent
\begin{tabular*}{\linewidth}{lc@{\extracolsep{\fill}}r@{\extracolsep{0pt}}}
\ifthenelse{\boolean{pdflatex}}
{\vspace*{-2.7cm}\mbox{\!\!\!\includegraphics[width=.14\textwidth]{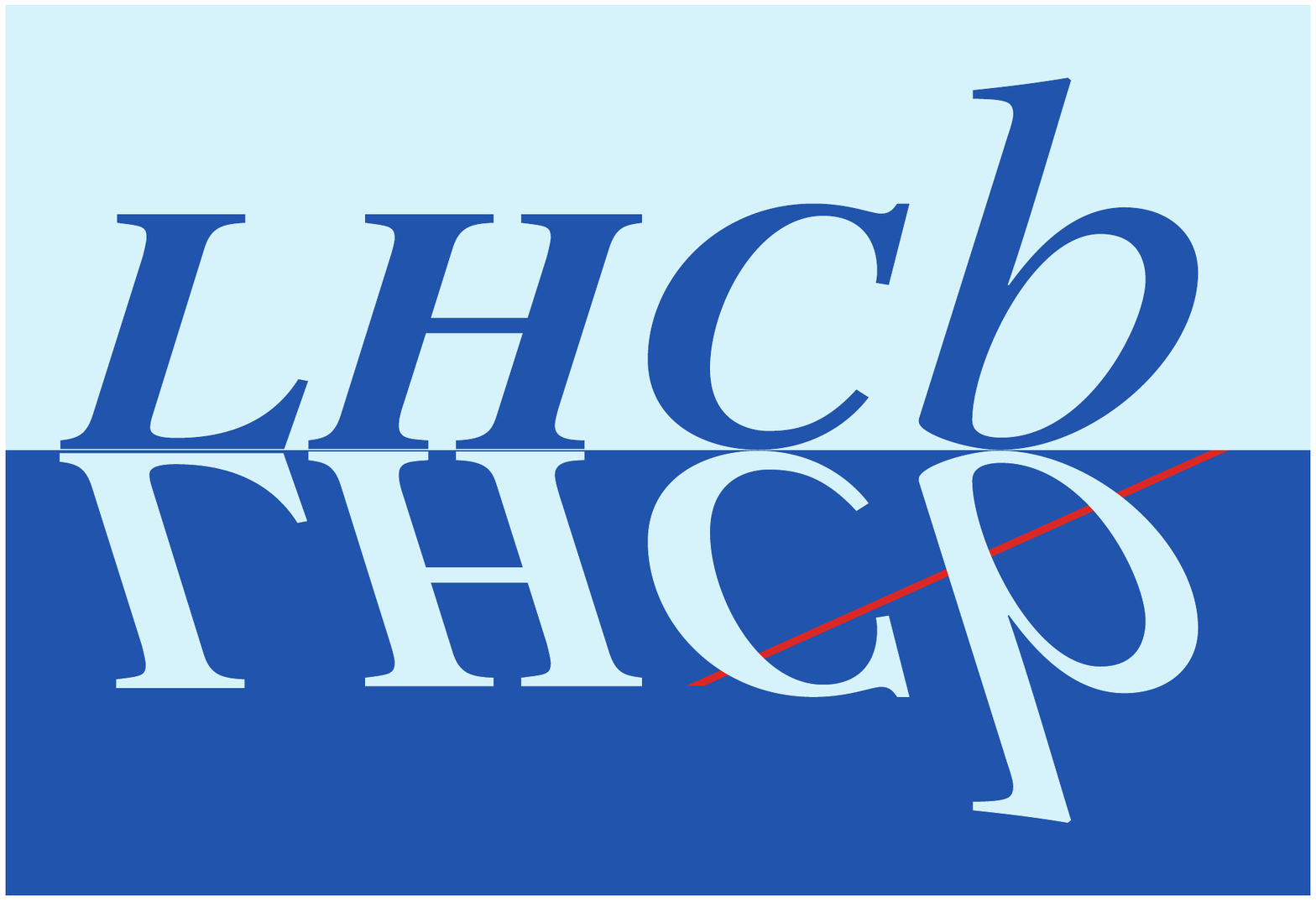}} & &}%
{\vspace*{-1.2cm}\mbox{\!\!\!\includegraphics[width=.12\textwidth]{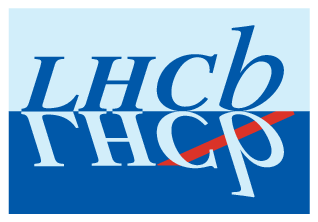}} & &}%
\\
 & & CERN-PH-EP-2015-219 \\  
 & & LHCb-PAPER-2015-035 \\  
 & & 1 September 2015 \\ 
 & & \\
\end{tabular*}

\vspace*{2.0cm}

{\bf\boldmath\huge
\begin{center}
First measurement of the differential branching fraction and~\CP~asymmetry of the $\Bpmpimm$ decay
\end{center}
}

\vspace*{1.0cm}

\begin{center}
The LHCb collaboration\footnote{Authors are listed at the end of this paper.}
\end{center}

\vspace{\fill}

\begin{abstract}
  \noindent
The differential branching fraction with respect to the dimuon invariant mass squared, and the \CP asymmetry of the \Bpmpimm decay are measured for the first time. The CKM matrix elements $|\Vtd|$ and $|\Vts|$, and the ratio $|\Vtd/\Vts|$ are determined.
The analysis is performed using proton-proton collision data corresponding to an integrated luminosity of 3.0\invfb, collected by the LHCb experiment at centre-of-mass energies of 7 and 8\tev. 
The total branching fraction and \CP asymmetry of \Bpmpimm decays are measured to be
\begin{align}
\BF(\Bpmpimm) &= (1.83 \pm 0.24 \pm 0.05) \times 10^{-8}\,\,\,\mathrm{and} \nonumber\\
\acp(\Bpmpimm) &= -0.11 \pm 0.12 \pm 0.01\,, \nonumber
\end{align}
where the first uncertainties are statistical and the second are systematic. These are the most precise measurements of these observables to date, and they are compatible with the predictions of the Standard Model.
\end{abstract}

\vspace*{1.0cm}

\begin{center}
Submitted to JHEP
\end{center}

\vspace{\fill}

{\footnotesize 
\centerline{\copyright~CERN on behalf of the \lhcb collaboration, licence \href{http://creativecommons.org/licenses/by/4.0/}{CC-BY-4.0}.}}
\vspace*{2mm}

\end{titlepage}


\newpage
\setcounter{page}{2}
\mbox{~}

\cleardoublepage


\renewcommand{\thefootnote}{\arabic{footnote}}
\setcounter{footnote}{0}



\pagestyle{plain} 
\setcounter{page}{1}
\pagenumbering{arabic}


\section{Introduction}
\label{sec:Introduction}

\begin{figure}[H]
\centering
\includegraphics[width=0.9\textwidth]{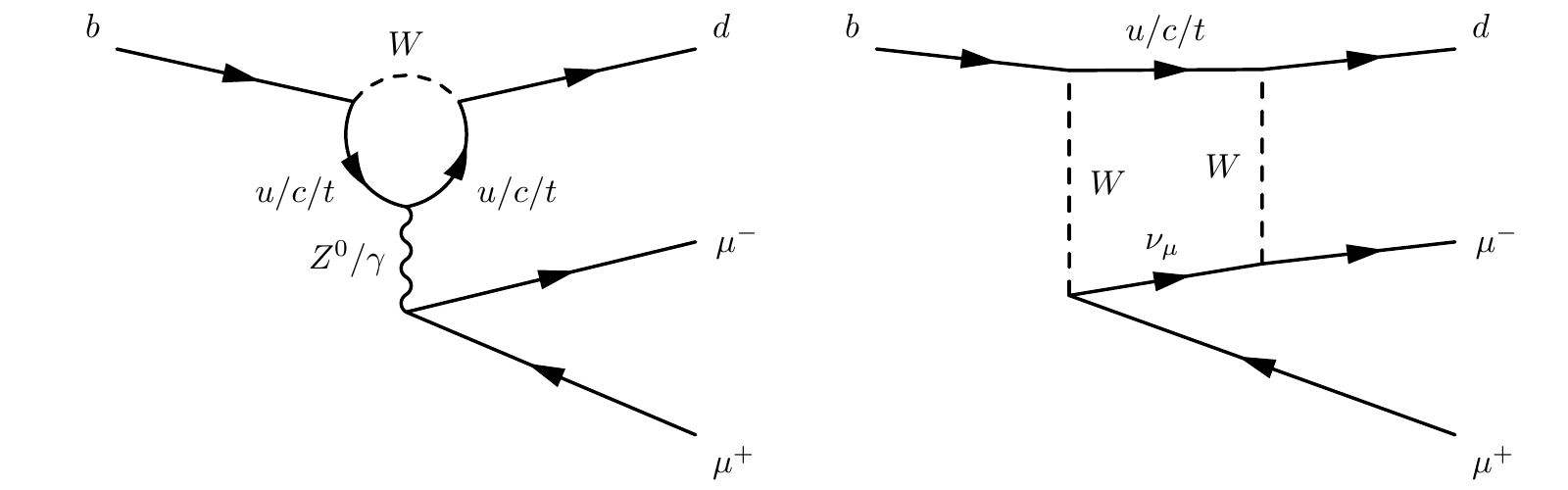}
\caption{Feynman diagrams of the penguin and box loop contributions to the \btodll process. 
~\label{fig:feyn}}
\end{figure}

The decay \Bupimm is a $\bquark\to\dquark$ flavour-changing neutral-current process, which is suppressed in the Standard Model  (SM).\footnote{Unless explicitly stated, the inclusion of charge-conjugate processes is implied.}
The suppression arises since the \btodll transition proceeds only through amplitudes involving the electroweak loop (penguin and box) diagrams shown in Fig.~\ref{fig:feyn}. 
\noindent In the SM, the top quark contribution dominates the loops, and an additional suppression occurs through the factor $\Vtd$ from the Cabbibo-Kobayashi-Maskawa (CKM) matrix. 
The decay is therefore sensitive to the presence of new particles that are predicted to exist in extensions of the SM, particularly in models where the flavour structure differs from that of the SM~\cite{Ali:2013zfa,Bartsch:2010qp,Wang:2007sp,Li:2014uha,Hou:2014dza,Hambrock:2015wka,Bailey:2015aba}. 
The ratio of CKM matrix elements $|\Vtd/\Vts|$ has been measured~\cite{PDG2014} via \Bd and \Bs mixing processes \cite{PhysRevLett.97.242003,LHCb-PAPER-2013-036} and $\bquark\to\squark(\dquark)\g$ decays~\cite{PhysRevD.82.051101}; it can also be determined from a measurement of the ratio of the branching fractions of the \Bupimm decay to the more precisely measured \BuKmm decay~\cite{LHCb-PAPER-2014-006}. Such ratios are also sensitive to the flavour structure of physics beyond the SM.

The \CP asymmetry of \Bpmpimm is defined as the relative difference between the decay widths, $\Gamma$, of the two charge conjugate modes,
\begin{align}
\acp &\equiv \frac{\Gamma(\Bmtopimumu) - \Gamma(\Bptopimumu) }{\Gamma(\Bmtopimumu) + \Gamma(\Bptopimumu)}\,.\label{eqn:acpdef}
\end{align}
The \CP asymmetry is predicted to be non-zero due to interference between amplitudes that are proportional to the CKM matrix elements involved in the \Bupimm decay, namely \Vub\Vuds and \Vtb\Vtds. 
Recent predictions for the \CP asymmetry are given in Ref.~\cite{Hambrock:2015wka}.
The \Bupimm decay was first observed by the \lhcb collaboration~\cite{LHCb-PAPER-2012-020} and the total branching fraction was measured to be
\begin{align}
\BF(\Bupimm) = (2.3\pm0.6\stat\pm0.1\syst)\times 10^{-8}\,.\nonumber
\end{align}

This paper describes measurements of the differential branching fraction and \CP asymmetry of the \Bpmpimm decay. The differential branching fraction is measured in bins of dilepton invariant mass squared, \qsq, and normalised to \BuJpsiK decays. These measurements are performed through fits to the invariant mass distributions. The branching fraction and the ratio of the branching fractions ${\BF(\Bupimm)/\BF(\BuKmm)}$ are used to determine the CKM matrix elements $|\Vtd|$ and $|\Vts|$, and the ratio $|\Vtd/\Vts|$, respectively. 
The measurements are based on 3.0\invfb of $pp$ collision data recorded using the \lhcb detector at centre-of-mass energies of 7\tev and 8\tev.

\section{Detector and simulation}
\label{sec:det}

The \lhcb detector~\cite{Alves:2008zz,LHCb-DP-2014-002} is a single-arm forward spectrometer covering the \mbox{pseudorapidity} range $2<\eta <5$, designed for the study of particles containing \bquark or \cquark quarks. 
The detector includes a high-precision tracking system consisting of a silicon-strip vertex detector surrounding the $pp$~interaction region, a large-area silicon-strip detector located upstream of a dipole magnet with a bending power of about $4{\rm\,Tm}$, and three stations of silicon-strip detectors and straw drift tubes placed downstream of the magnet.
The tracking system provides a measurement of the momentum,~\ptot, of charged particles with a relative uncertainty that varies from 0.5\% at low momentum to 1.0\% at 200\gevc.
The minimum distance of a track to a primary vertex, the impact parameter, is measured with a resolution of $(15+29/\pt)\mum$, where \pt is the component of the momentum transverse to the beam, in\,\gevc. 
The magnetic field polarity is inverted with a period of several weeks during data taking, which allows the charge asymmetries due to the detector geometry to be determined.

The different types of charged hadrons are distinguished using information from two ring-imaging Cherenkov detectors. 
Photons, electrons and hadrons are identified by a calorimeter system consisting of scintillating-pad and preshower detectors, an electromagnetic calorimeter and a hadronic calorimeter. 
Muons are identified by a system composed of alternating layers of iron and multiwire proportional chambers.
The online event selection is performed by a trigger, which consists of a hardware stage, based on information from the calorimeter and muon systems, followed by a software stage, which reconstructs the full event.

Samples of simulated \Bupimm, \BuKmm and \BuJpsiK decays are produced from $pp$ collisions generated using \pythia~\cite{Sjostrand:2006za,*Sjostrand:2007gs} with a specific \lhcb configuration~\cite{LHCb-PROC-2010-056}.  
Decays of hadronic particles are described by \evtgen~\cite{Lange:2001uf}, in which final-state radiation is generated using \photos~\cite{Golonka:2005pn}. 
The interaction of the generated particles with the detector, and its response, are implemented using the \geant toolkit~\cite{Allison:2006ve, *Agostinelli:2002hh} as described in Ref.~\cite{LHCb-PROC-2011-006}.
The simulated events are reweighted to account for known differences relative to the data in the transverse momentum spectrum of the \Bu meson and the detector occupancy of the event.

\section{Event selection}
\label{sec:sel}

Events are required to satisfy a hardware trigger, which selects muons with $\pt>1.48\gevc$ in the 7\tev data and $\pt>1.76\gevc$ in the 8\tev data. 
In the subsequent software trigger, at least one of the final-state particles is required to have both $\pt>0.8\gevc$ and impact parameter greater than $100\mum$ with respect to all primary $pp$ interaction vertices~(PVs) in the event. 
Finally, the tracks of at least two of the final-state particles are required to form a vertex that is significantly displaced from the PVs, and a multivariate algorithm is used to identify secondary vertices that are consistent with the decay of a \bquark hadron~\cite{LHCb-DP-2014-002}.

Candidates are formed from pairs of well-reconstructed oppositely-charged tracks identified as muons, combined with an additional track that is identified as either a charged pion or a charged kaon for \Bupimm or \BuKmm decays, respectively.
Each track is required to have a good fit quality, a low probability of overlapping with any other track, $\pt>300\mevc$ and to be inconsistent with originating from any PV.
Candidates are required to have a good quality vertex fit and to be consistent with originating from a PV with the candidate's momentum vector aligned with the direction between the primary and secondary vertices.

Separation of the signal decay from combinatorial background is achieved using a multivariate classifier. 
A boosted decision tree~(BDT)~\cite{Breiman,AdaBoost} is trained using supervised learning with ten-fold cross validation~\cite{Stone:1974} to achieve an unbiased classifier response.
The background sample used to train the BDT consists of data from the upper sideband of the \pimumu invariant mass distribution in the region greater than 5500\mevcc; the \Bupimm signal sample is obtained from the simulation. As no particle identification information is used in the classifier, it can be applied to both the pion and kaon modes. 
The features of the data that are used to classify the \pimumu candidate as signal- or background-like are the properties of the pion and muon tracks, and properties of the \pimumu candidate. 
For the pion and muon tracks, the features used are the transverse momentum of the tracks, the impact parameter of the track, and the track quality.
For the \pimumu candidate, the features used are the angle between its momentum vector and the direction vector between the primary vertex and the secondary vertex, and its flight distance, transverse momentum, and vertex quality. Two isolation variables~\cite{LHCB-PAPER-2015-025} and the absolute difference in momentum between each of the muons are also used in the classifier. 

The output of the multivariate classifier and the particle identification requirements are simultaneously optimised to maximise signal significance. Pseudo-datasets were constructed from simulated signal events and combinatorial background events taken from the upper mass sideband of data. Trial BDT and particle identification cuts were applied and an expected misidentified-kaon component added to the pseudo-datasets. Wilks' theorem~\cite{Wilks:1938dza} was used to determine a signal significance from fits to the pseudo-dataset, the value of which was passed to a maximisation algorithm that could vary the trial cut values.
The classifier and particle identification cut values used to separate signal and background decays are chosen at the point of highest significance. Operating at this point, the classifier has a combinatorial background rejection of 99.8\%, whilst retaining 66.9\% of signal events, and each event contains only a single candidate. 
As the classifier separates \Bu decays from combinatorial background, relatively pure samples of \BuKmm and \BuJpsiK events are also obtained using the same classifier requirements, when requiring a positively identified kaon.

The charmonium resonances are removed from the samples of ${\Bupimm}$ and ${\BuKmm}$ candidates by vetoing the regions ${8.0<\qsq<11.0\gevgevcccc}$ and ${12.5<\qsq<15.0\gevgevcccc}$. 
There are several other \bquark-hadron decays that could mimic the \Bupimm signal.
Decays such as \decay{\Bu}{\pip\pim\pip} and \BuJpsiK, where there is double hadron-muon misidentification, are excluded from the \Bupimm dataset by muon identification criteria and the expected number of background events is found to be negligible.
Partially reconstructed decays such as \decay{\Bd}{\Kstarz(\Kp\pim)\mumu}, \decay{\Bd}{\KS(\pip\pim)\mumu} and \decay{\Bd}{\rho(\pip\pim)\mumu}, where a kaon or a pion is missed, 
may satisfy the selection; however, simulation indicates that such events have a reconstructed mass that lies more than 100\mevcc below the measured \Bu mass. Therefore, such background events do not affect the signal yield extraction.

There are two types of semileptonic decays that feature as backgrounds, ${\decay{\Bu}{\Dzb(\Kp\mun\neumb)\pip}}$ decays with kaon-muon misidentification, and the double semileptonic decay ${\decay{\Bu}{\Dzb(h^{+}\mun\neumb)\mup\neum}}$, where $h^+$ can be a pion or kaon.
The former decay is suppressed by requiring the \mup to have a low probability of being a kaon.
The latter decay has the same final state as the signal and cannot be completely removed by the selection.
However, the distribution of double semileptonic decays as a function of the \pimumu invariant mass varies smoothly, and can be modelled well in the fit from which the signal yield is extracted. 
The pion-kaon separation is not completely efficient: 6\% of \BuKmm events are selected as \Bupimm events, and are modelled as a specific background.
The normalisation sample of \BuJpsiK candidates is selected using the dilepton invariant-mass region around the \jpsi mass, \ie $3096\pm50\mevcc$. 
To remove much of the contribution from partially reconstructed decays, whilst keeping enough information to determine any effect on the signal, the \pimumu invariant-mass range $5040<m(\pimumu)<6000\mevcc$ is used to extract the signal yield.

\section{Event yields}
\label{sec:yields}

The yields of \Bupimm, \BuKmm and \BuJpsiK candidates are extracted by performing simultaneous, extended, unbinned maximum-likelihood fits to the invariant mass distributions $m(\pimumu)$ and $m(\kmumu)$ of the selected candidates. 
The total model for the invariant mass distribution is composed of a signal model, a combinatorial background model, a model to describe partially reconstructed \bquark-meson decays and a model to describe \bquark-hadron decays with misidentified final-state particles.
The signal model is an empirical function that consists of two Gaussian functions with power-law tails on both sides~\cite{Skwarnicki:1986xj}, and the same parameters are used for the ${\Bupimm}$, ${\BuKmm}$, and ${\BuJpsiK}$ decay modes. 
The model for the combinatorial background is described by a separate exponential function for each decay.
In the ${\Bupimm}$ data sample, the misidentified \BuKmm decays where a kaon has been misidentified as a pion, are described by a single Gaussian function with a power-law tail on the lower-mass side. 
The yield of misidentified \BuKmm decays is constrained using the measured branching fraction~\cite{LHCb-PAPER-2014-006} and the observed pion-kaon misidentification efficiency.
The mass distribution of the misidentified \BuKmm candidates is obtained by fitting the invariant mass distribution of \BuJpsiK candidates, where the kaon is required to have the pion mass, 
and which has been corrected to account for differences in the particle identification efficiencies that arise from the differing kinematics.
The partially reconstructed \Bu decays in the \BuKmm and the \BuJpsiK data are described by an empirical function, which consists of a rising exponential function that makes a smooth transition to a Gaussian function.
This description allows the mixture of partially reconstructed \bquark-hadron decays to be limited to less than the maximum physical value of the \Bu mass minus the pion mass, with a Gaussian resolution-smearing effect.

The partially reconstructed \bquark-hadron decays in the \Bupimm sample are separated into three explicit components. 
Firstly, the double semileptonic decay ${\BuDzpimumu}$ is included, as this is an irreducible background that ends at the \Bu mass. This is modelled by a falling exponential function that makes a smooth transition to a Gaussian function at high mass, where the parameters are fixed from a fit to simulated events. The yield of this component is left to vary in the fit. 
Secondly, the decays \Burhomumu and \Bdrhomumu are estimated to contribute a total of $34\pm7$ events to the data, from the measured branching fraction of \Bdrhomumu~\cite{LHCb-PAPER-2014-063} and assuming isospin invariance. 
Lastly, the decay \Bsfzmumu is estimated to contribute $10\pm2$ events to the data, also below the \Bu mass.
Each of these decays is modelled by a separate kernel-estimation probability density function (PDF) with a shape taken from simulated events reconstructed under the \pimumu hypothesis. The yield of each of these decays has a Gaussian constraint applied with a central value and width set to the expected yield and its uncertainty.

\begin{figure}[tbp]
\centering
\includegraphics[width=0.45\textwidth]{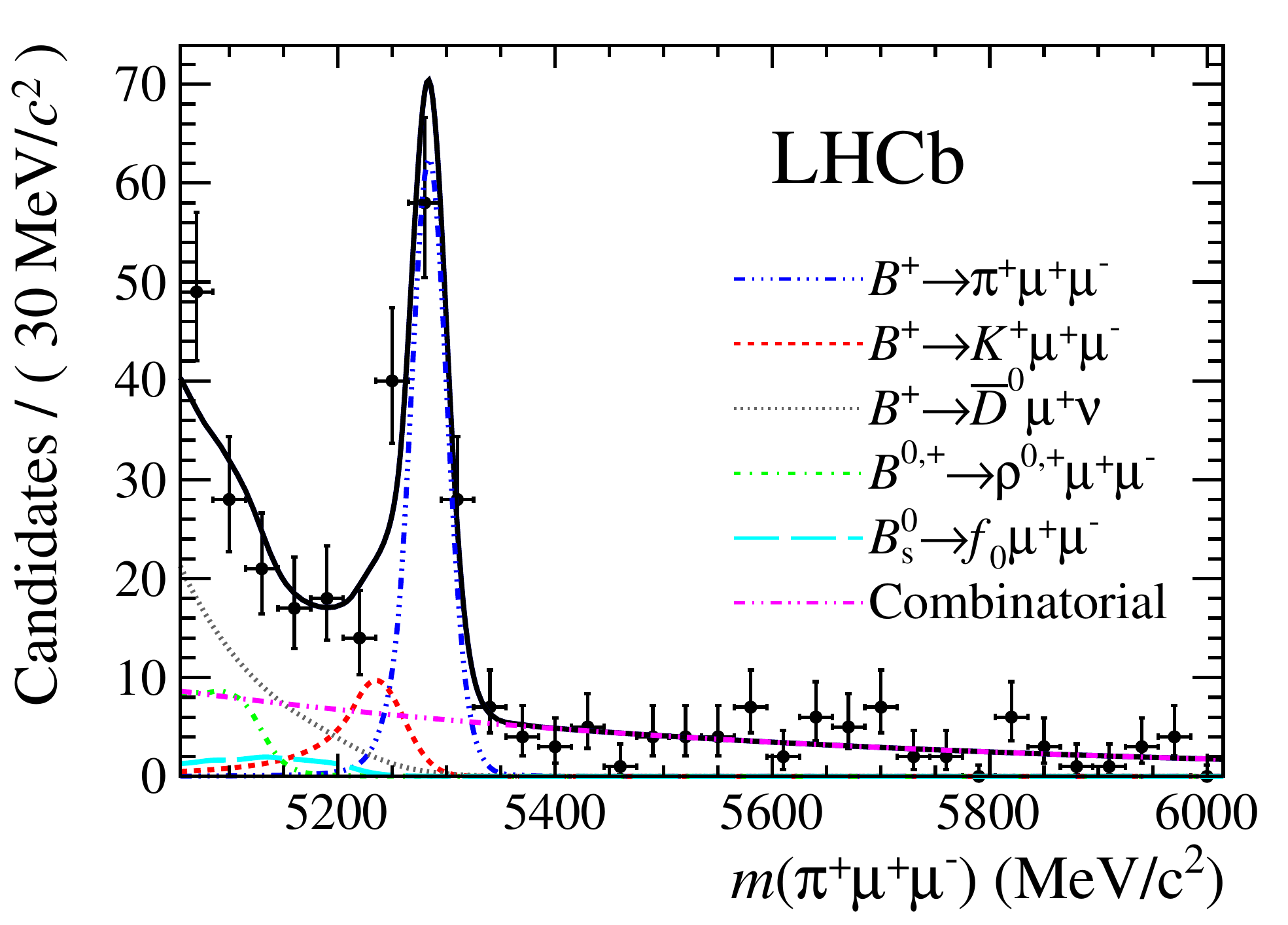}
\includegraphics[width=0.45\textwidth]{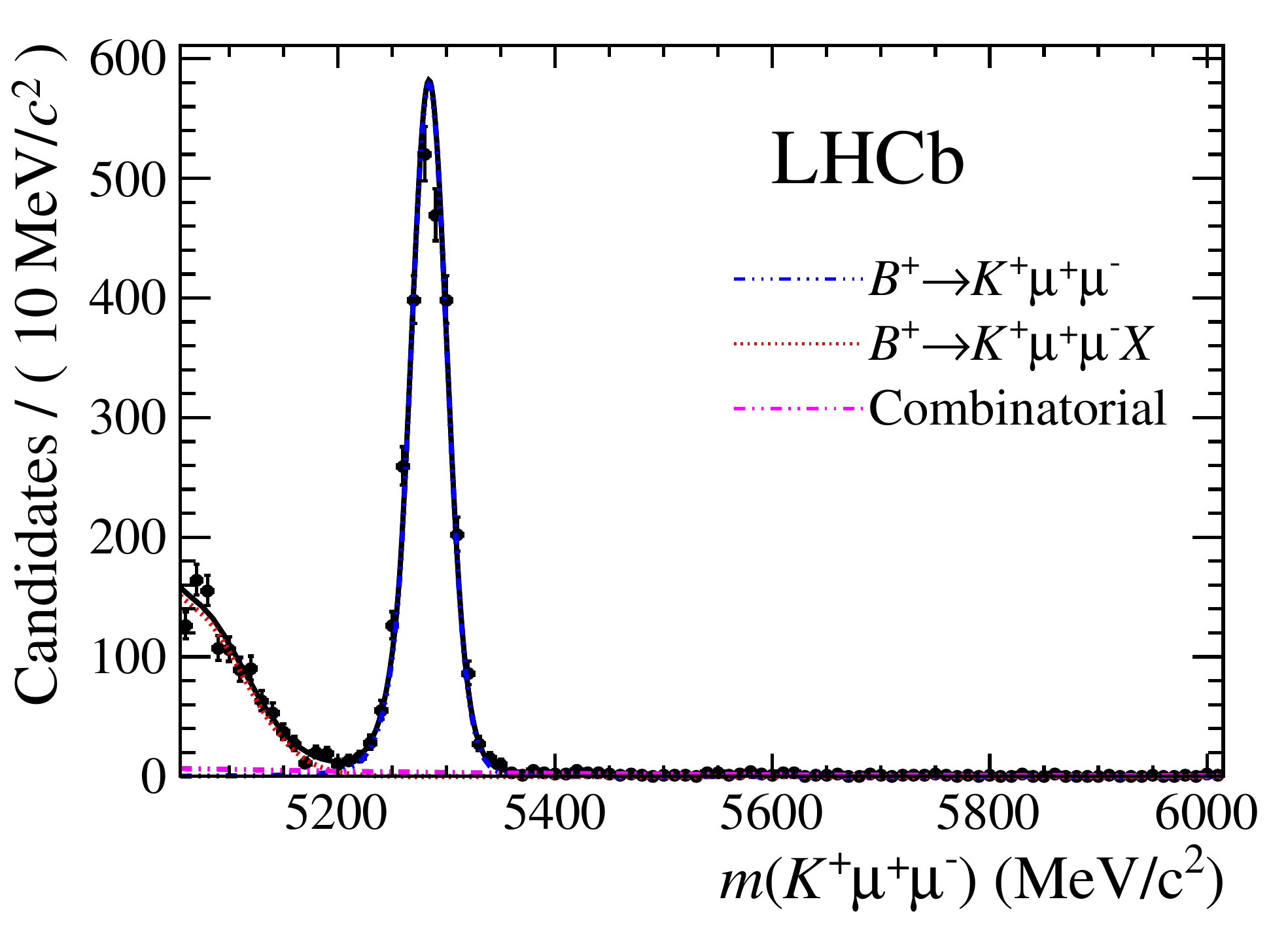}
\caption{The fit to the invariant mass distribution of (left) selected \Bupimm candidates and (right) selected \BuKmm candidates, with the total model and separate components as described in the legend.~\label{fig:fits}}
\end{figure}

\begin{figure}[tbp]
\centering
\includegraphics[width=0.45\textwidth]{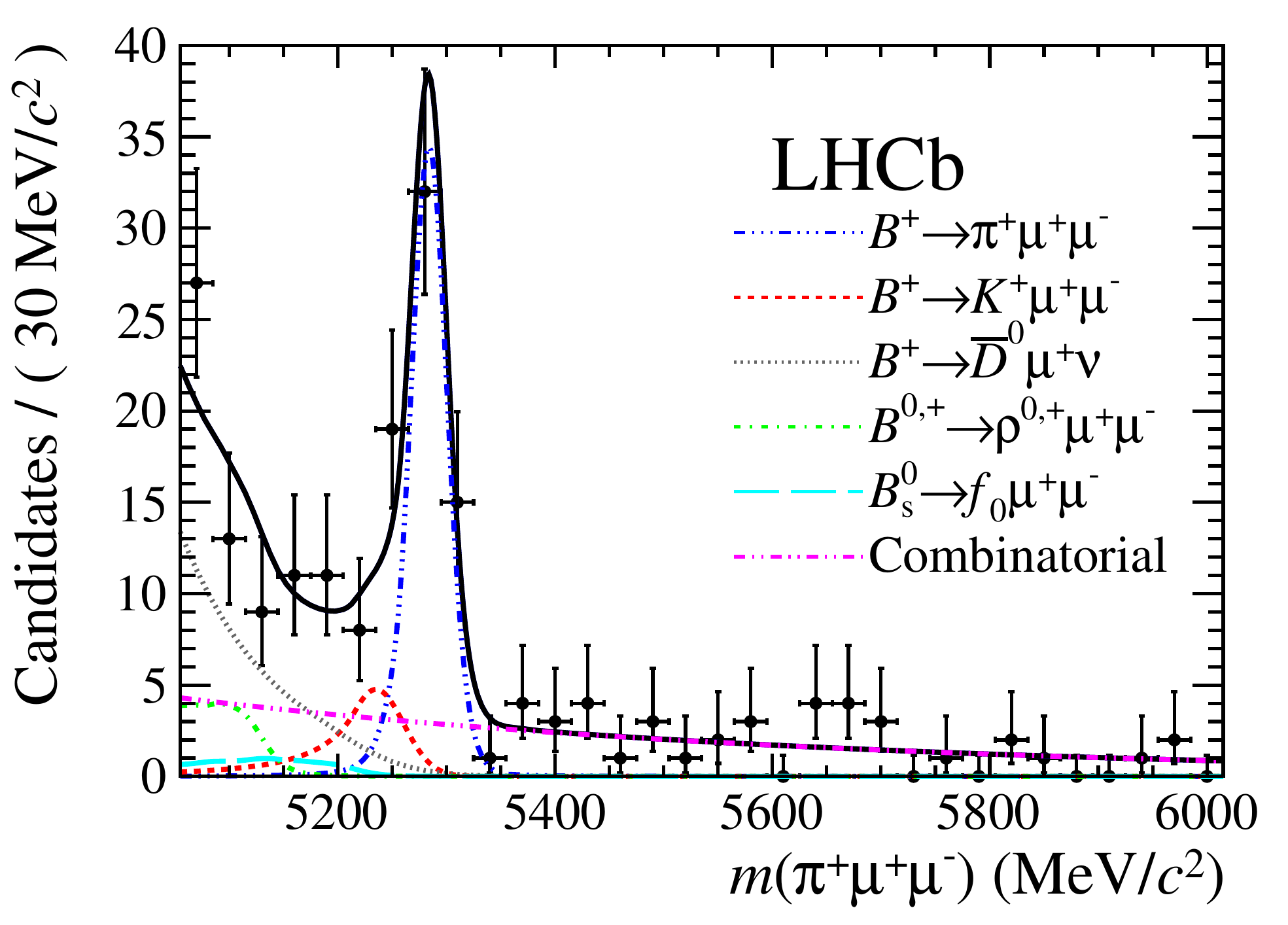}
\includegraphics[width=0.45\textwidth]{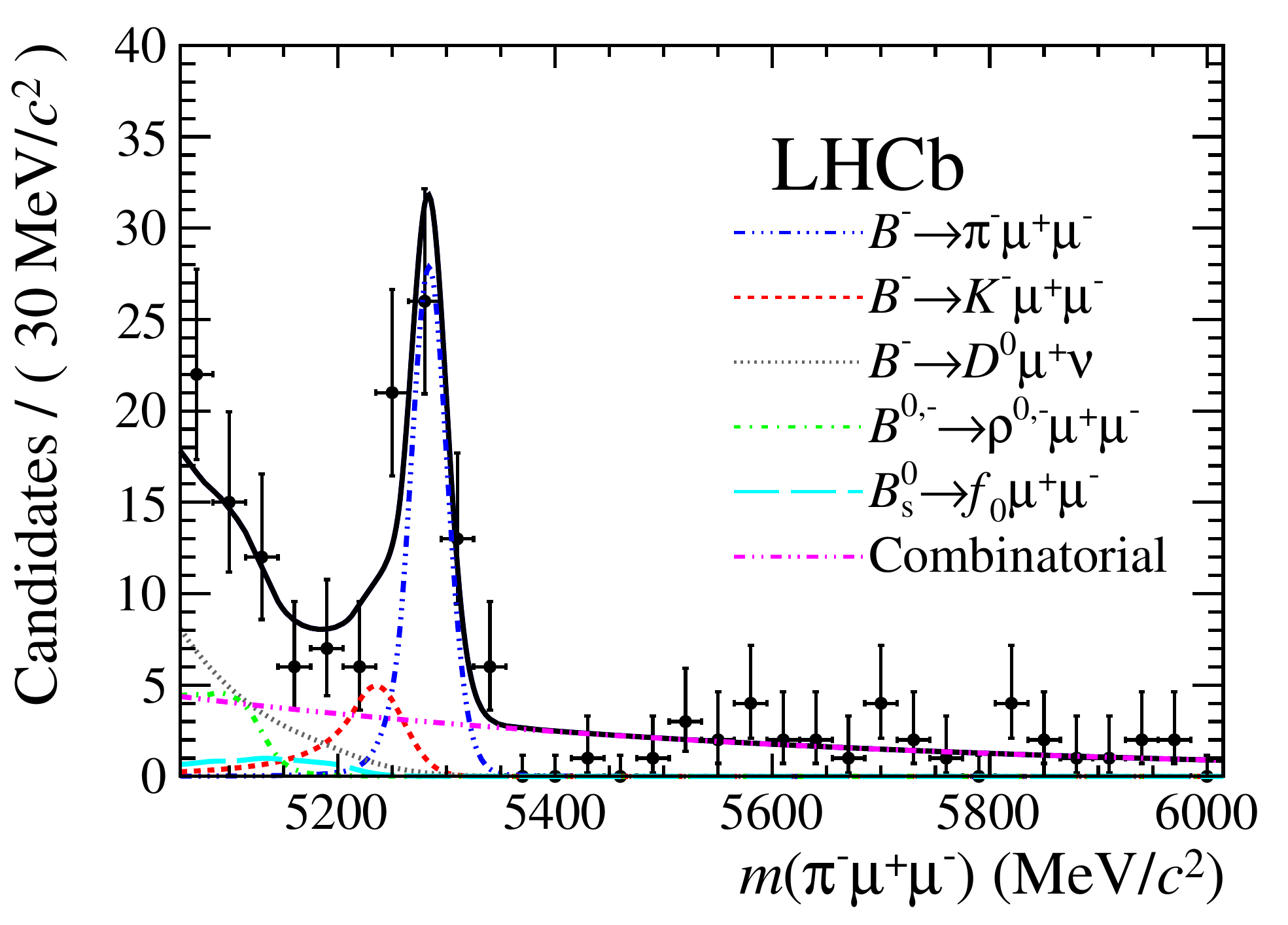}
\caption{The fit to the invariant mass distribution of (left) selected \Bupimm candidates and (right) selected \Bmpimm candidates, with the total model and separate components as described in the legend.~\label{fig:pol}}
\end{figure}

The invariant mass distributions of selected \pimumu and \kmumu candidates are shown in Fig.~\ref{fig:fits}, along with the total fitted model, signal component, and each background component.
The fit gives yields of $94\pm12$ \Bupimm, $2922\pm55$ \BuKmm, and ${(609.5\pm0.8)\times 10^3\,\BuJpsiK}$ candidates, where the uncertainties are statistical.
The yield of \Bupimm in each \qsq bin is given in Table~\ref{tbl:yields}. The ratio of CKM matrix elements is determined in the theoretically favourable \cite{Ali:2013zfa} bins $1.0 < \qsq < 6.0\gevgevcccc$ (low-\qsq) and $15.0 < \qsq < 22.0\gevgevcccc$ (high-\qsq). The \BuKmm yields are $879\pm30$ in the low-\qsq bin and $793\pm28$ in the high-\qsq bin. 
\setlength\extrarowheight{4pt}
\begin{table}[tbp]
\centering
\caption{The yields of \Bupimm decays in bins of dilepton invariant mass squared, with statistical uncertainties.~\label{tbl:yields}}
\begin{tabular}{r@{\;--\;}l|r@{\;}c@{\;}l}
\multicolumn{2}{c|}{\qsq bin (\gevgevcccc )}& \multicolumn{3}{c}{\Bupimm}     \\ 
\hline
\hspace{18pt}0.1 &	2.0&	\hspace{10pt}$22.5$ & $^+_-$ & $^{5.5}_{4.8}$  \\
\hspace{18pt}2.0 &	4.0&	$7.5$  & $^+_-$ & $^{4.9}_{4.0}$               \\
\hspace{18pt}4.0 &	6.0&	$11.1$ & $^+_-$ & $^{4.2}_{3.5}$               \\
\hspace{18pt}6.0 &	8.0&	$9.5$  & $\pm$  & $ 3.9$                       \\
\hspace{18pt}11.0 &	12.5&	$10.5$ & $\pm$  & $ 3.7$                       \\
\hspace{18pt}15.0 &	17.0&	$9.7$  & $\pm$  & $ 3.3$                       \\
\hspace{18pt}17.0 &	19.0&	$6.2$  & $\pm$  & $ 2.9$                       \\
\hspace{18pt}19.0 &	22.0&	$7.8$  & $\pm$  & $3.4$                        \\
\hspace{18pt}22.0 & 25.0&	$2.3$  & $^+_-$ & $^{2.1}_{1.5}$               \\
\hspace{18pt}0.0 &	25.0&	$93.6$ & $\pm$  & $ 11.5$                      \\
\hline
\hspace{18pt}1.0 &	6.0&	$28.8$ & $^+_-$ & $^{6.7}_{6.2}$               \\
\hspace{18pt}15.0 &	22.0&	$24.1$ & $^+_-$ & $^{6.0}_{5.2}$               \\
\end{tabular}
\end{table}
The results of a simultaneous fit to the invariant mass distribution of \Bupimm and \Bmpimm candidates are shown in Fig.~\ref{fig:pol} 
\begin{table}[htbp]
\centering
\caption{The measured total yield from the simultaneous fit to the charge separated data, and the inferred yields of \Bupimm and \Bmpimm decays.~\label{tbl:pol}}
\begin{tabular}{c|c|c}
 $\mathcal{N}(\Bpmpimm)$ & $\mathcal{N}(\Bppimm)$ & $\mathcal{N}(\Bmpimm)$\\  
\hline
92.7 $\pm$ 11.5 & $51.7 \pm 8.3$ & $41.1 \pm 7.9$ \\ 
\end{tabular}
\end{table}
and the measured yields are given in Table~\ref{tbl:pol}. The small difference in total signal yield between this fit and that given in Table~\ref{tbl:yields} is due to the systematic effect of separating the background distributions by charge. Consistent results are obtained from datasets split between the two magnet polarities.
\newpage
The choice of models used for the partially reconstructed backgrounds, the semileptonic backgrounds, the misidentified \kmumu background, and the combinatorial background could all contribute as potential sources of systematic uncertainty.
The dependence of the fitted yields on these models is assessed by replacing the relevant component with an alternative model, as follows, and evaluating the change in yield in simulation studies and in the fits to data. The largest change in yield is assigned as the systematic uncertainty. 
Changing the models for the \Burhomumu and \Bdrhomumu decays to an exponential function with a Gaussian high-mass endpoint contributes 0.6\% uncertainty to the measured \Bupimm yield, and using an analogous shape for the \Bsfzmumu decays contributes 0.7\%. The parameters of the models are fixed to values obtained from a fit to the simulation.
The systematic uncertainty of the model used for the semileptonic backgrounds is evaluated by allowing the exponent in the model to vary within the uncertainties produced by a fit to the simulation. This change contributes 0.3\% uncertainty to the measured \Bupimm yield. There is a negligible contribution from altering the model of the misidentified decays or combinatorial background, and from changing the upper mass end-point of the fit range from 6000\mevcc to either 5500 or 7000\mevcc.

\section{Results}
\label{sec:res}

\subsection{Differential branching fraction}

The differential branching fraction of \Bupimm in a bin of width ${\Delta\qsq}$ is calculated relative to the normalisation channel \BuJpsiK as
\begin{align}
\frac{\deriv\BF(\Bupimm)}{\dqsq} = \frac{\mathcal{N}_{\Bupimm}}{\epsilon_{\Bupimm}}\!\times\!\frac{\epsilon_{\BuJpsiK}}{\mathcal{N}_{\BuJpsiK}}\!\times\!\frac{\BF(\BuJpsiK) }{\Delta\qsq}\,,
\end{align}
where $\mathcal{N}$ is the event yield, $\epsilon$ is the total efficiency to select the decay, both of which are functions of \qsq, and $\BF(\BuJpsiK)=(1.05\pm0.05)\times10^{-3}$ is the measured branching fraction of the normalisation channel, with $\BF(\jpsi\to\mup\mun) = (5.961\pm0.033)\%$~\cite{PDG2014}.

The total efficiency to select the candidates for the decays considered is computed from the product of the efficiencies to trigger, reconstruct and select the final-state particles and the \Bu candidate.
This includes the geometrical acceptance of the \lhcb detector and the efficiencies of the trigger and selection algorithms.
These efficiencies are calculated using a combination of simulated signal events and data-driven methods. 
The use of the ratio of efficiencies of the decay modes ensures that many of the possible sources of systematic uncertainty largely cancel.
The efficiency of the trigger depends on the kinematics of the muons, and this dependence contributes a source of systematic uncertainty relative to the signal yield at the level of $2\%$.
The dependence of the particle identification efficiency on the kinematic distributions contributes a systematic uncertainty of $<0.1\%$ for the muons, $2\%$ for the pions and $<0.1\%$ for the kaons. These uncertainties are evaluated by varying the binning of the kinematic variables, and include a contribution from the size of the calibration samples used.
The calculation of the BDT efficiency is affected by small differences between the simulation and data. The dependence of the signal yield on these differences is assessed using the \BuJpsiK and \BuJpsipi decays. The relatively large yield allows precise comparisons of data and simulation.
The impact of using simulation to calculate the efficiency of the BDT is assessed using the observed differences between data and simulation in the normalisation channel; a systematic uncertainty of $1.4\%$ is assigned.

\begin{figure}[tbp]
\centering
\includegraphics[width=0.7\textwidth]{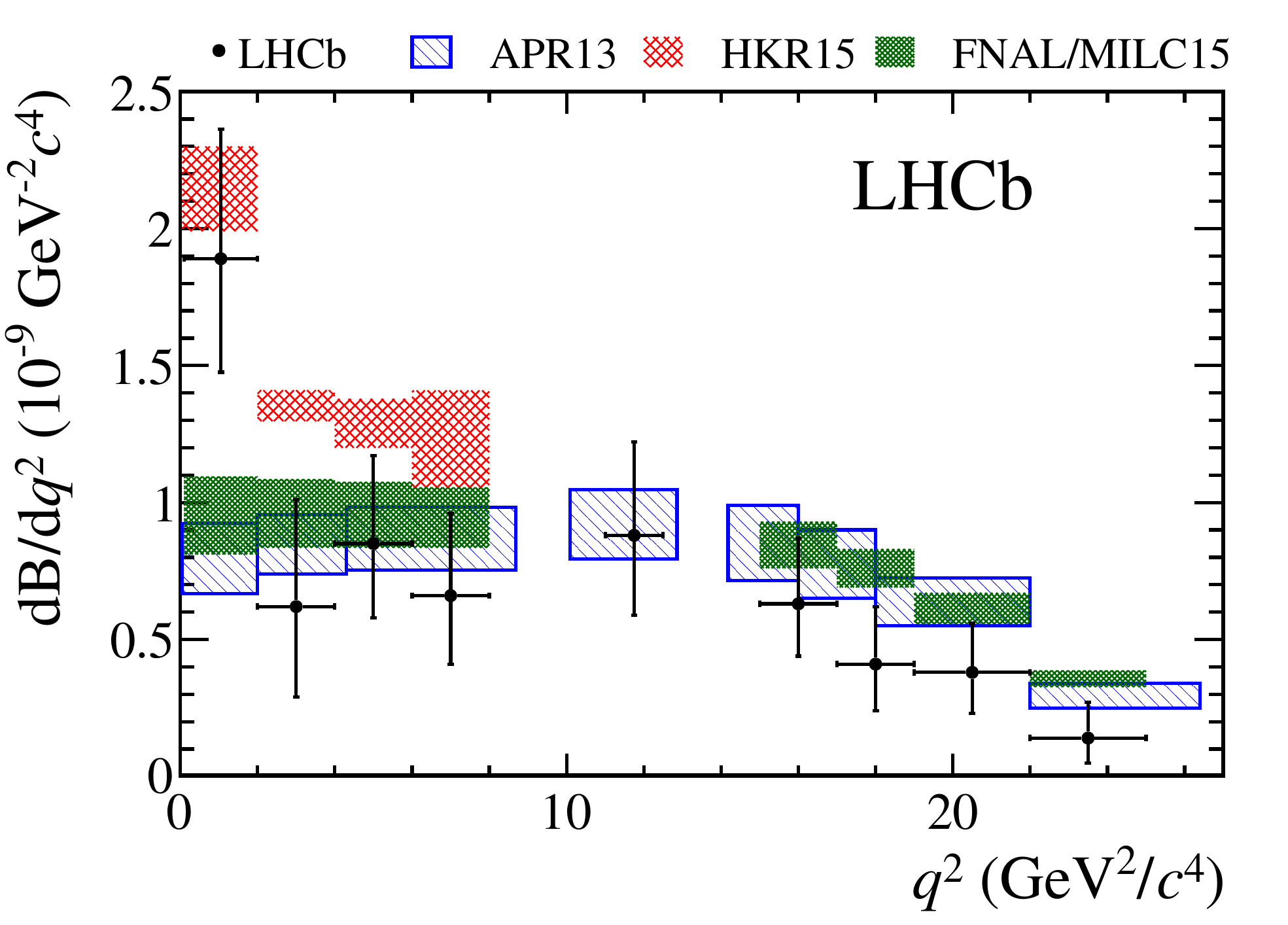}
\caption{The differential branching fraction of \Bupimm in bins of dilepton invariant mass squared, \qsq, compared to SM predictions taken from Refs.~\cite{Ali:2013zfa} (APR13), \cite{Hambrock:2015wka} (HKR15) and from lattice QCD calculations~\cite{Bailey:2015aba} (FNAL/MILC15).~\label{fig:dbr}}
\end{figure}

\begin{figure}[tbp]
\centering
\includegraphics[width=0.7\textwidth]{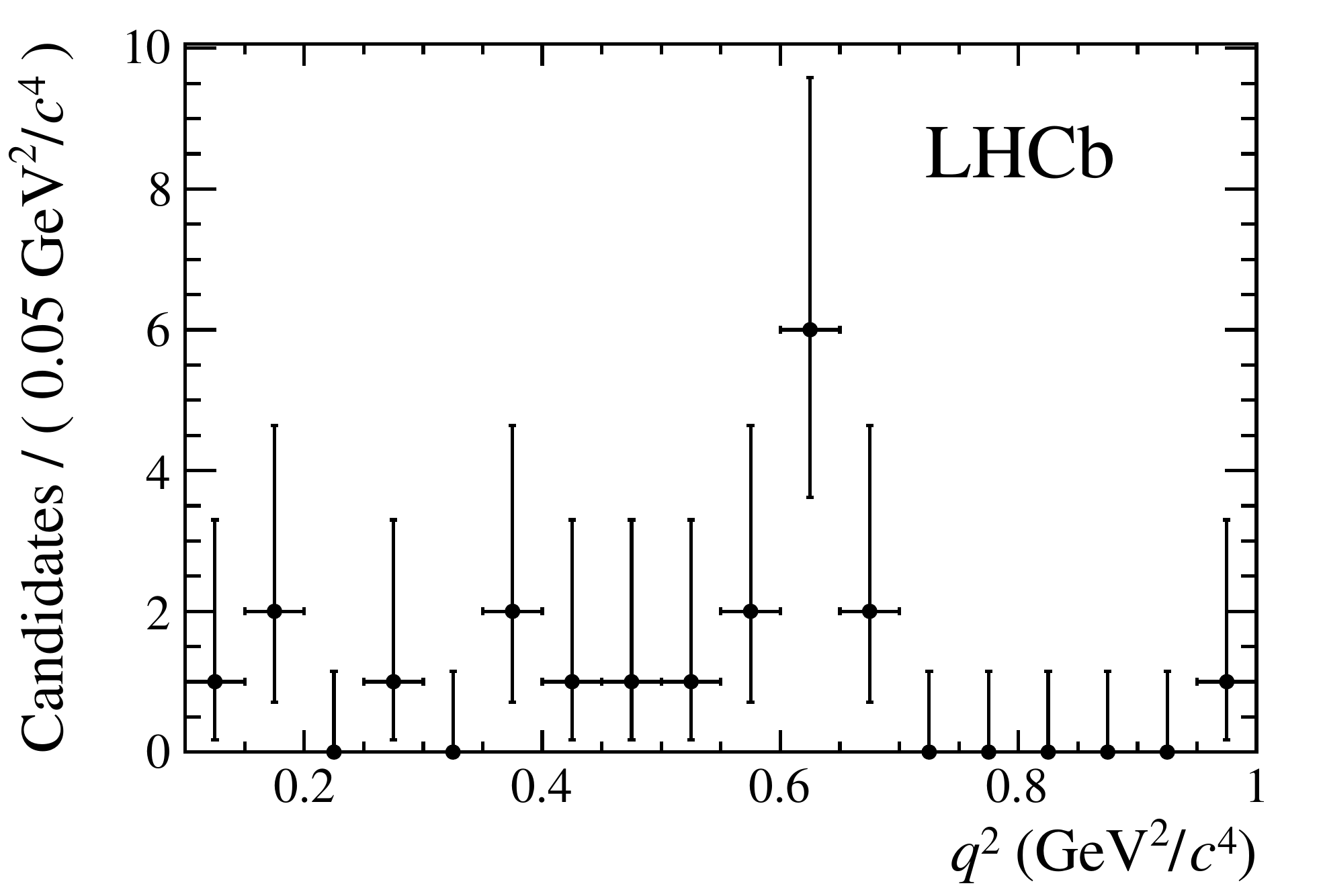}
\caption{The \qsq spectrum of \Bupimm candidates in the region 0.1 -- 1.0\gevgevcccc in a $\pm 50$~\mev window around the nominal \Bu mass, showing a peaking structure at 0.6\gevgevcccc that is in the region of the \rhoz and $\omega$ masses squared.~\label{fig:rho}}
\end{figure}

The measured values of the differential branching fraction are shown in Fig.~\ref{fig:dbr} and given in Table~\ref{tbl:dbr}. The branching fraction agrees with SM predictions from Refs.~\cite{Ali:2013zfa,Hambrock:2015wka}, although agreement in the lowest-\qsq bin is only achieved when contributions from low-\qsq resonances are taken into account, as in Ref.~\cite{Hambrock:2015wka}. The \qsq spectrum of candidates below 1\gevgevcccc in a $\pm 50$~\mev window around the nominal \Bu mass is shown in Fig.~\ref{fig:rho}, with hints of a peaking structure in the vicinity of the \rhoz and $\omega$ masses. The total branching fraction is computed from the integral over the measured bins multiplied by a scaling factor to account for the regions of \qsq not measured in this analysis. This factor is taken from simulation to be $1.333 \pm 0.004$, where the uncertainty combines the statistical and systematic uncertainties evaluated by using two different form factor models. The total branching fraction is therefore 
\begin{align}\BF(\Bupimm) = (1.83 \pm 0.24 \stat \pm 0.05\syst) \times 10^{-8}\,. \nonumber\end{align}
The ratio of branching fractions of $\BF(\Bupimm)$ to $\BF(\BuKmm)$ in the region $1.0 < \qsq < 6.0\gevgevcccc$ is
\begin{align}\frac{\BF(\Bupimm)}{\BF(\BuKmm)} =  0.038 \pm 0.009\stat \pm 0.001\syst\,, \nonumber\end{align}
and in the region $15.0 < \qsq < 22.0\gevgevcccc$ is
\begin{align}\frac{\BF(\Bupimm)}{\BF(\BuKmm)} =  0.037 \pm 0.008\stat \pm 0.001\syst\,. \nonumber\end{align}
These results are the most precise measurements of these quantities to date.

\begin{table}[htbp]
\centering
\caption{The results for the differential branching fraction for \Bupimm in bins of \qsq. 
The first uncertainties are statistical and the second are systematic.~\label{tbl:dbr}}
\setlength\extrarowheight{4pt}
\begin{tabular}{c|c}
\qsq bin (\gevgevcccc)   & $\frac{\deriv\BR}{\deriv\qsq}(\Bupimm)$ ($10^{-9}\gev^{-2} c^4$)\\[0.5em]
\hline
0.1 -- 2.0 &$ 1.89\,^{+0.47}_{-0.41} \pm 0.06 $ \\
2.0 -- 4.0 &$ 0.62\,^{+0.39}_{-0.33} \pm 0.02 $ \\
4.0 -- 6.0 &$ 0.85\,^{+0.32}_{-0.27} \pm 0.02 $ \\
6.0 -- 8.0 &$ 0.66\,^{+0.30}_{-0.25} \pm 0.02 $ \\
11.0 -- 12.5 &$ 0.88\,^{+0.34}_{-0.29} \pm 0.03 $ \\
15.0 -- 17.0 &$ 0.63\,^{+0.24}_{-0.19} \pm 0.02 $ \\
17.0 -- 19.0 &$ 0.41\,^{+0.21}_{-0.17} \pm 0.01 $ \\
19.0 -- 22.0 &$ 0.38\,^{+0.18}_{-0.15} \pm 0.01 $ \\
22.0 -- 25.0 &$ 0.14\,^{+0.13}_{-0.09} \pm 0.01 $ \\
\hline
1.0 -- 6.0 &$ 0.91\,^{+0.21}_{-0.20} \pm 0.03 $ \\
15.0 -- 22.0 &$ 0.47\,^{+0.12}_{-0.10} \pm 0.01 $ \\
\end{tabular}
\end{table}

\subsection{CKM matrix elements}

The ratio of CKM matrix elements $|\Vtd/\Vts|$ can be calculated from the ratio of branching fractions, $\BF(\Bupimm)/\BF(\BuKmm)$, and is given in terms of measured quantities
\begin{align}
|\Vtd/\Vts|^2 & = \frac{\BF(\Bupimm)}{\BF(\BuKmm)} \times \frac{\int F_K\dqsq}{\int F_\pi\dqsq}\,
\end{align}
where $F_{\pi(K)}$ is the combination of form factor, Wilson coefficients and phase space factor for the $\Bu\to\pion(\kaon)$ decay.
The values of $\int F_{\pi, K}\dqsq$ are calculated using the EOS package~\cite{Bobeth:2010wg}, with $\Bp\to\pip$ form factors taken from Refs.~\cite{Bourrely:2008za,Imsong:2014oqa} and $\Bp\to\Kp$ form factors taken from Ref.~\cite{Ball:2004ye}.
The EOS package is a framework for calculating observables, with uncertainties, in semileptonic \bquark-quark decays for both SM and new physics parameters. 
In order to take into account the correlations between the theory inputs for the matrix element ratio calculation, the EOS package is used to produce a PDF as a function of the \Bupimm and \BuKmm branching fractions in each of the relevant \qsq bins by Monte Carlo sampling of the theory nuisance parameters. A $\chi^2$ minimisation is performed to determine $|\Vtd/\Vts|$, taking into account the data and this PDF, and the theory nuisance parameters are free to vary. The data are treated as uncorrelated between the two \qsq bins, but the full correlation between the theory parameters is accounted for. The value of the CKM matrix element ratio is determined to be
\begin{align}
\left|\frac{\Vtd}{\Vts}\right| = 0.24^{+0.05}_{-0.04}\,,\nonumber
\end{align}
where the uncertainty is the combination of the experimental (statistical and systematic), and theoretical uncertainties. Both contributions are approximately equal, and neither follows a Gaussian distribution. 
This is the most precise determination of $|\Vtd/\Vts|$ in a decay that includes both penguin and box diagrams.

Additionally, the values of $|\Vtd|$ and $|\Vts|$ can be calculated via 
\begin{align}
|\Vtd|^2 &= \frac{\BR(\Bupimm) }{\int F_{\pi}\dqsq}\,\,\,\mathrm{and}\\
|\Vts|^2 &= \frac{\BR(\BuKmm) }{\int F_{\PK} \dqsq}\,,
\end{align}
where EOS is used to compute the theoretical input. Combining the results from the high- and low-\qsq bins gives
\begin{align}
\left|\Vtd\right| &= 7.2^{+0.9}_{-0.8} \times 10^{-3}\nonumber\,\,\,\mathrm{and} \\[0.5em]
\left|\Vts\right| &= 3.2^{+0.4}_{-0.4} \times 10^{-2}\,,\nonumber
\end{align}
where the uncertainties are due to both the branching fraction measurements and the theory nuisance parameters. As the $|\Vtd/\Vts|$ determination uses both the \Bupimm and \BuKmm branching fraction measurements, the theory nuisance parameters take different values to those in the separate $|\Vtd|$ and $|\Vts|$ determinations, where only one of the branching fractions is used. The ratio of $|\Vtd|$ and $|\Vts|$ is therefore not identical to the measurement of $|\Vtd/\Vts|$ given above.
 The uncertainty on $|\Vtd|$ has approximately equal contributions from experimental and theoretical uncertainties, while the uncertainty on $|\Vts|$ is dominated by the theoretical uncertainty.

\subsection{\textbf{\textit{C}}$\!$\textbf{\textit{P}} asymmetry}

The \CP asymmetry of \Bpmpimm, as defined by Eq.~\ref{eqn:acpdef}, can be computed from the raw yield asymmetry, 
\begin{align}
\araw &\equiv \frac{\mathcal{N}(\Bmtopimumu) - \mathcal{N}(\Bptopimumu) }{\mathcal{N}(\Bmtopimumu) + \mathcal{N}(\Bptopimumu)}\,,
\end{align}
where $\mathcal{N}$ is the signal yield for the given decay-mode. This raw asymmetry is corrected for the production asymmetry of the \Bpm mesons and the detection asymmetry of the decay products, under the approximation 
\begin{align}
\acp(\Bpmpimm) = \araw - \ap - \adet\,,
\end{align}
where \ap is the \Bpm-meson production asymmetry, and \adet is the detector asymmetry for the pions and muons.

The production asymmetry of \Bp and \Bm mesons at \lhcb has been measured to be ${(-0.6\pm0.6)\%}$ using the \BuJpsiK decay~\cite{LHCb-PAPER-2014-053}. 
The momentum spectrum differences between the \BuJpsiK and \Bupimm decays are found to have a negligible impact on this asymmetry. 
The charge asymmetry of the \lhcb detector for \pip and \pim has been measured in \Dstarpm decays~\cite{LHCb-PAPER-2012-009} to be $\varepsilon_\pip/\varepsilon_\pim = 0.9914\pm0.0040$ and $\varepsilon_\pip/\varepsilon_\pim = 1.0045\pm0.0034$ for the two magnet polarities.
These efficiency ratios give detector asymmetries of $(-0.43\pm0.20)\%$ and $(0.22\pm0.17)\%$ for the two magnet polarities, where the differences in the momentum spectrum are accounted for in bins of momentum, transverse momentum and azimuthal angle.
The relative tracking efficiency of differently charged pions is consistent with unity when averaged over the the two magnet polarities~\cite{LHCb-PAPER-2012-009}.
The pion identification asymmetry is derived using $\Dz\to\Km\pip$ decays and is calculated to be less than 0.087\% when momentum spectrum differences are accounted for. 
Additional effects from the production and detection asymmetries are negligible and do not contribute to the final systematic uncertainty.

The raw \CP asymmetry, \araw, of the \Bpmpimm candidates is measured to be $-0.11 \pm 0.12$.
The value of \acp for \Bpmpimm is calculated to be
\begin{align}\acp(\Bpmpimm) = -0.11 \pm 0.12\stat \pm 0.01\syst\,,\nonumber\end{align}
which is consistent with a recent SM prediction~\cite{Hambrock:2015wka}.

\section{Summary}
\label{sec:conc}

A measurement of the differential branching fraction of the decay \Bupimm has been presented, and is found to be consistent with SM predictions, and to have a possible contribution from $\Bu\to\rhoz(\omega)\pip$ decays. The \CP asymmetry of the decay has been measured and is consistent with a recent SM prediction~\cite{Hambrock:2015wka}. The values for the CKM matrix elements $|\Vtd|$ and $|\Vts|$, and the ratio $|\Vtd/\Vts|$ have also been determined, and are in agreement with previous measurements. These results constitute the most precise measurements to date of a \btodll transition and supersede those of Ref.~\cite{LHCb-PAPER-2012-020}.

\section*{Acknowledgements}

\noindent The authors would like to thank Danny van Dyk for his assistance in using the EOS software package and Alexander Khodjamirian for advice on calculating the CKM matrix elements. 
We express our gratitude to our colleagues in the CERN
accelerator departments for the excellent performance of the LHC. We
thank the technical and administrative staff at the LHCb
institutes. We acknowledge support from CERN and from the national
agencies: CAPES, CNPq, FAPERJ and FINEP (Brazil); NSFC (China);
CNRS/IN2P3 (France); BMBF, DFG, HGF and MPG (Germany); INFN (Italy); 
FOM and NWO (The Netherlands); MNiSW and NCN (Poland); MEN/IFA (Romania); 
MinES and FANO (Russia); MinECo (Spain); SNSF and SER (Switzerland); 
NASU (Ukraine); STFC (United Kingdom); NSF (USA).
The Tier1 computing centres are supported by IN2P3 (France), KIT and BMBF 
(Germany), INFN (Italy), NWO and SURF (The Netherlands), PIC (Spain), GridPP 
(United Kingdom).
We are indebted to the communities behind the multiple open 
source software packages on which we depend. We are also thankful for the 
computing resources and the access to software R\&D tools provided by Yandex LLC (Russia).
Individual groups or members have received support from 
EPLANET, Marie Sk\l{}odowska-Curie Actions and ERC (European Union), 
Conseil g\'{e}n\'{e}ral de Haute-Savoie, Labex ENIGMASS and OCEVU, 
R\'{e}gion Auvergne (France), RFBR (Russia), XuntaGal and GENCAT (Spain), Royal Society and Royal
Commission for the Exhibition of 1851 (United Kingdom).

\addcontentsline{toc}{section}{References}
\setboolean{inbibliography}{true}
\bibliographystyle{LHCb}
\bibliography{main,LHCb-PAPER,LHCb-CONF,LHCb-DP,LHCb-TDR,theo}

\ifx\mcitethebibliography\mciteundefinedmacro
\PackageError{LHCb.bst}{mciteplus.sty has not been loaded}
{This bibstyle requires the use of the mciteplus package.}\fi
\providecommand{\href}[2]{#2}
\begin{mcitethebibliography}{10}
\mciteSetBstSublistMode{n}
\mciteSetBstMaxWidthForm{subitem}{\alph{mcitesubitemcount})}
\mciteSetBstSublistLabelBeginEnd{\mcitemaxwidthsubitemform\space}
{\relax}{\relax}

\bibitem{Ali:2013zfa}
A.~Ali, A.~Y. Parkhomenko, and A.~V. Rusov,
  \ifthenelse{\boolean{articletitles}}{\emph{{Precise calculation of the
  dilepton invariant-mass spectrum and the decay rate in $B^\pm \to \pi^\pm
  \mu^+ \mu^-$ in the SM}},
  }{}\href{http://dx.doi.org/10.1103/PhysRevD.89.094021}{Phys.\ Rev.\
  \textbf{D89} (2014) 094021}, \href{http://arxiv.org/abs/1312.2523}{{\tt
  arXiv:1312.2523}}\relax
\mciteBstWouldAddEndPuncttrue
\mciteSetBstMidEndSepPunct{\mcitedefaultmidpunct}
{\mcitedefaultendpunct}{\mcitedefaultseppunct}\relax
\EndOfBibitem
\bibitem{Bartsch:2010qp}
M.~Bartsch, M.~Beylich, G.~Buchalla, and D.-N. Gao,
  \ifthenelse{\boolean{articletitles}}{\emph{{Precision flavour physics with
  $B\to K \nu \neub$ and $B \to K\ellell$}},
  }{}\href{http://dx.doi.org/10.1088/1126-6708/2009/11/011}{JHEP \textbf{0911}
  (2009) 011}, \href{http://arxiv.org/abs/0909.1512}{{\tt
  arXiv:0909.1512}}\relax
\mciteBstWouldAddEndPuncttrue
\mciteSetBstMidEndSepPunct{\mcitedefaultmidpunct}
{\mcitedefaultendpunct}{\mcitedefaultseppunct}\relax
\EndOfBibitem
\bibitem{Wang:2007sp}
J.-J. Wang, R.-M. Wang, Y.-G. Xu, and Y.-D. Yang,
  \ifthenelse{\boolean{articletitles}}{\emph{{The rare decays $B^+_u \to \pi^+
  \ellell, \rho^+ \ellell$ and $B^0_d \to \ellell$ in the R-parity violating
  supersymmetry}},
  }{}\href{http://dx.doi.org/10.1103/PhysRevD.77.014017}{Phys.\ Rev.\
  \textbf{D77} (2008) 014017}, \href{http://arxiv.org/abs/0711.0321}{{\tt
  arXiv:0711.0321}}\relax
\mciteBstWouldAddEndPuncttrue
\mciteSetBstMidEndSepPunct{\mcitedefaultmidpunct}
{\mcitedefaultendpunct}{\mcitedefaultseppunct}\relax
\EndOfBibitem
\bibitem{Li:2014uha}
Z.-H. Li, Z.-G. Si, Y.~Wang, and N.~Zhu,
  \ifthenelse{\boolean{articletitles}}{\emph{{$B\to \pi\ell^{+}\ell^{-}$ decays
  revisited in the Standard Model}},
  }{}\href{http://arxiv.org/abs/1411.0466}{{\tt arXiv:1411.0466}}\relax
\mciteBstWouldAddEndPuncttrue
\mciteSetBstMidEndSepPunct{\mcitedefaultmidpunct}
{\mcitedefaultendpunct}{\mcitedefaultseppunct}\relax
\EndOfBibitem
\bibitem{Hou:2014dza}
W.-S. Hou, M.~Kohda, and F.~Xu,
  \ifthenelse{\boolean{articletitles}}{\emph{{Rates and asymmetries of
  $B\to\pi\ellell$ decays}},
  }{}\href{http://dx.doi.org/10.1103/PhysRevD.90.013002}{Phys.\ Rev.\
  \textbf{D90} (2014) 013002}, \href{http://arxiv.org/abs/1403.7410}{{\tt
  arXiv:1403.7410}}\relax
\mciteBstWouldAddEndPuncttrue
\mciteSetBstMidEndSepPunct{\mcitedefaultmidpunct}
{\mcitedefaultendpunct}{\mcitedefaultseppunct}\relax
\EndOfBibitem
\bibitem{Hambrock:2015wka}
C.~Hambrock, A.~Khodjamirian, and A.~Rusov,
  \ifthenelse{\boolean{articletitles}}{\emph{{Hadronic effects and observables
  in $B\to \pi\ell^+\ell^-$ decay at large recoil}},
  }{}\href{http://arxiv.org/abs/1506.07760}{{\tt arXiv:1506.07760}}\relax
\mciteBstWouldAddEndPuncttrue
\mciteSetBstMidEndSepPunct{\mcitedefaultmidpunct}
{\mcitedefaultendpunct}{\mcitedefaultseppunct}\relax
\EndOfBibitem
\bibitem{Bailey:2015aba}
J.~A. Bailey {\em et~al.},
  \ifthenelse{\boolean{articletitles}}{\emph{{$B\to\pi\ell\ell$ form factors
  for new-physics searches from lattice QCD}},
  }{}\href{http://arxiv.org/abs/1507.01618}{{\tt arXiv:1507.01618}}\relax
\mciteBstWouldAddEndPuncttrue
\mciteSetBstMidEndSepPunct{\mcitedefaultmidpunct}
{\mcitedefaultendpunct}{\mcitedefaultseppunct}\relax
\EndOfBibitem
\bibitem{PDG2014}
Particle Data Group, K.~A. Olive {\em et~al.},
  \ifthenelse{\boolean{articletitles}}{\emph{{\href{http://pdg.lbl.gov/}{Review
  of particle physics}}},
  }{}\href{http://dx.doi.org/10.1088/1674-1137/38/9/090001}{Chin.\ Phys.\
  \textbf{C38} (2014) 090001}\relax
\mciteBstWouldAddEndPuncttrue
\mciteSetBstMidEndSepPunct{\mcitedefaultmidpunct}
{\mcitedefaultendpunct}{\mcitedefaultseppunct}\relax
\EndOfBibitem
\bibitem{PhysRevLett.97.242003}
CDF collaboration, A.~Abulencia {\em et~al.},
  \ifthenelse{\boolean{articletitles}}{\emph{Observation of
  ${B}^0_s$--$\overline{B}^0_s$ oscillations},
  }{}\href{http://dx.doi.org/10.1103/PhysRevLett.97.242003}{Phys.\ Rev.\ Lett.\
   \textbf{97} (2006) 242003}\relax
\mciteBstWouldAddEndPuncttrue
\mciteSetBstMidEndSepPunct{\mcitedefaultmidpunct}
{\mcitedefaultendpunct}{\mcitedefaultseppunct}\relax
\EndOfBibitem
\bibitem{LHCb-PAPER-2013-036}
LHCb collaboration, R.~Aaij {\em et~al.},
  \ifthenelse{\boolean{articletitles}}{\emph{{Observation of
  $B^0_s$--$\overline{B}^0_s$ mixing and measurement of mixing frequencies
  using semileptonic $B$ decays}},
  }{}\href{http://dx.doi.org/10.1140/epjc/s10052-013-2655-8}{Eur.\ Phys.\ J.\
  \textbf{C73} (2013) 2655}, \href{http://arxiv.org/abs/1308.1302}{{\tt
  arXiv:1308.1302}}\relax
\mciteBstWouldAddEndPuncttrue
\mciteSetBstMidEndSepPunct{\mcitedefaultmidpunct}
{\mcitedefaultendpunct}{\mcitedefaultseppunct}\relax
\EndOfBibitem
\bibitem{PhysRevD.82.051101}
\babar collaboration, P.~del Amo~Sanchez {\em et~al.},
  \ifthenelse{\boolean{articletitles}}{\emph{Study of ${B}\rightarrow
  {X}\gamma$ decays and determination of $|{V}_{td}/{V}_{ts}|$},
  }{}\href{http://dx.doi.org/10.1103/PhysRevD.82.051101}{Phys.\ Rev.\ D
  \textbf{82} (2010) 051101}, \href{http://arxiv.org/abs/1005.4087}{{\tt
  arXiv:1005.4087}}\relax
\mciteBstWouldAddEndPuncttrue
\mciteSetBstMidEndSepPunct{\mcitedefaultmidpunct}
{\mcitedefaultendpunct}{\mcitedefaultseppunct}\relax
\EndOfBibitem
\bibitem{LHCb-PAPER-2014-006}
LHCb collaboration, R.~Aaij {\em et~al.},
  \ifthenelse{\boolean{articletitles}}{\emph{{Differential branching fractions
  and isospin asymmetries of $B \to K^{(*)}\mu^+\mu^-$ decays}},
  }{}\href{http://dx.doi.org/10.1007/JHEP06(2014)133}{JHEP \textbf{06} (2014)
  133}, \href{http://arxiv.org/abs/1403.8044}{{\tt arXiv:1403.8044}}\relax
\mciteBstWouldAddEndPuncttrue
\mciteSetBstMidEndSepPunct{\mcitedefaultmidpunct}
{\mcitedefaultendpunct}{\mcitedefaultseppunct}\relax
\EndOfBibitem
\bibitem{LHCb-PAPER-2012-020}
LHCb collaboration, R.~Aaij {\em et~al.},
  \ifthenelse{\boolean{articletitles}}{\emph{{First observation of the decay
  $B^+ \to \pi^+ \mu^+\mu^-$}},
  }{}\href{http://dx.doi.org/10.1007/JHEP12(2012)125}{JHEP \textbf{12} (2012)
  125}, \href{http://arxiv.org/abs/1210.2645}{{\tt arXiv:1210.2645}}\relax
\mciteBstWouldAddEndPuncttrue
\mciteSetBstMidEndSepPunct{\mcitedefaultmidpunct}
{\mcitedefaultendpunct}{\mcitedefaultseppunct}\relax
\EndOfBibitem
\bibitem{Alves:2008zz}
LHCb collaboration, A.~A. Alves~Jr.\ {\em et~al.},
  \ifthenelse{\boolean{articletitles}}{\emph{{The \lhcb detector at the LHC}},
  }{}\href{http://dx.doi.org/10.1088/1748-0221/3/08/S08005}{JINST \textbf{3}
  (2008) S08005}\relax
\mciteBstWouldAddEndPuncttrue
\mciteSetBstMidEndSepPunct{\mcitedefaultmidpunct}
{\mcitedefaultendpunct}{\mcitedefaultseppunct}\relax
\EndOfBibitem
\bibitem{LHCb-DP-2014-002}
LHCb collaboration, R.~Aaij {\em et~al.},
  \ifthenelse{\boolean{articletitles}}{\emph{{LHCb detector performance}},
  }{}\href{http://dx.doi.org/10.1142/S0217751X15300227}{Int.\ J.\ Mod.\ Phys.\
  \textbf{A30} (2015) 1530022}, \href{http://arxiv.org/abs/1412.6352}{{\tt
  arXiv:1412.6352}}\relax
\mciteBstWouldAddEndPuncttrue
\mciteSetBstMidEndSepPunct{\mcitedefaultmidpunct}
{\mcitedefaultendpunct}{\mcitedefaultseppunct}\relax
\EndOfBibitem
\bibitem{Sjostrand:2006za}
T.~Sj\"{o}strand, S.~Mrenna, and P.~Skands,
  \ifthenelse{\boolean{articletitles}}{\emph{{PYTHIA 6.4 physics and manual}},
  }{}\href{http://dx.doi.org/10.1088/1126-6708/2006/05/026}{JHEP \textbf{05}
  (2006) 026}, \href{http://arxiv.org/abs/hep-ph/0603175}{{\tt
  arXiv:hep-ph/0603175}}\relax
\mciteBstWouldAddEndPuncttrue
\mciteSetBstMidEndSepPunct{\mcitedefaultmidpunct}
{\mcitedefaultendpunct}{\mcitedefaultseppunct}\relax
\EndOfBibitem
\bibitem{Sjostrand:2007gs}
T.~Sj\"{o}strand, S.~Mrenna, and P.~Skands,
  \ifthenelse{\boolean{articletitles}}{\emph{{A brief introduction to PYTHIA
  8.1}}, }{}\href{http://dx.doi.org/10.1016/j.cpc.2008.01.036}{Comput.\ Phys.\
  Commun.\  \textbf{178} (2008) 852},
  \href{http://arxiv.org/abs/0710.3820}{{\tt arXiv:0710.3820}}\relax
\mciteBstWouldAddEndPuncttrue
\mciteSetBstMidEndSepPunct{\mcitedefaultmidpunct}
{\mcitedefaultendpunct}{\mcitedefaultseppunct}\relax
\EndOfBibitem
\bibitem{LHCb-PROC-2010-056}
I.~Belyaev {\em et~al.}, \ifthenelse{\boolean{articletitles}}{\emph{{Handling
  of the generation of primary events in Gauss, the LHCb simulation
  framework}}, }{}\href{http://dx.doi.org/10.1088/1742-6596/331/3/032047}{{J.\
  Phys.\ Conf.\ Ser.\ } \textbf{331} (2011) 032047}\relax
\mciteBstWouldAddEndPuncttrue
\mciteSetBstMidEndSepPunct{\mcitedefaultmidpunct}
{\mcitedefaultendpunct}{\mcitedefaultseppunct}\relax
\EndOfBibitem
\bibitem{Lange:2001uf}
D.~J. Lange, \ifthenelse{\boolean{articletitles}}{\emph{{The EvtGen particle
  decay simulation package}},
  }{}\href{http://dx.doi.org/10.1016/S0168-9002(01)00089-4}{Nucl.\ Instrum.\
  Meth.\  \textbf{A462} (2001) 152}\relax
\mciteBstWouldAddEndPuncttrue
\mciteSetBstMidEndSepPunct{\mcitedefaultmidpunct}
{\mcitedefaultendpunct}{\mcitedefaultseppunct}\relax
\EndOfBibitem
\bibitem{Golonka:2005pn}
P.~Golonka and Z.~Was, \ifthenelse{\boolean{articletitles}}{\emph{{PHOTOS Monte
  Carlo: A precision tool for QED corrections in $Z$ and $W$ decays}},
  }{}\href{http://dx.doi.org/10.1140/epjc/s2005-02396-4}{Eur.\ Phys.\ J.\
  \textbf{C45} (2006) 97}, \href{http://arxiv.org/abs/hep-ph/0506026}{{\tt
  arXiv:hep-ph/0506026}}\relax
\mciteBstWouldAddEndPuncttrue
\mciteSetBstMidEndSepPunct{\mcitedefaultmidpunct}
{\mcitedefaultendpunct}{\mcitedefaultseppunct}\relax
\EndOfBibitem
\bibitem{Allison:2006ve}
Geant4 collaboration, J.~Allison {\em et~al.},
  \ifthenelse{\boolean{articletitles}}{\emph{{Geant4 developments and
  applications}}, }{}\href{http://dx.doi.org/10.1109/TNS.2006.869826}{IEEE
  Trans.\ Nucl.\ Sci.\  \textbf{53} (2006) 270}\relax
\mciteBstWouldAddEndPuncttrue
\mciteSetBstMidEndSepPunct{\mcitedefaultmidpunct}
{\mcitedefaultendpunct}{\mcitedefaultseppunct}\relax
\EndOfBibitem
\bibitem{Agostinelli:2002hh}
Geant4 collaboration, S.~Agostinelli {\em et~al.},
  \ifthenelse{\boolean{articletitles}}{\emph{{Geant4: A simulation toolkit}},
  }{}\href{http://dx.doi.org/10.1016/S0168-9002(03)01368-8}{Nucl.\ Instrum.\
  Meth.\  \textbf{A506} (2003) 250}\relax
\mciteBstWouldAddEndPuncttrue
\mciteSetBstMidEndSepPunct{\mcitedefaultmidpunct}
{\mcitedefaultendpunct}{\mcitedefaultseppunct}\relax
\EndOfBibitem
\bibitem{LHCb-PROC-2011-006}
M.~Clemencic {\em et~al.}, \ifthenelse{\boolean{articletitles}}{\emph{{The
  \lhcb simulation application, Gauss: Design, evolution and experience}},
  }{}\href{http://dx.doi.org/10.1088/1742-6596/331/3/032023}{{J.\ Phys.\ Conf.\
  Ser.\ } \textbf{331} (2011) 032023}\relax
\mciteBstWouldAddEndPuncttrue
\mciteSetBstMidEndSepPunct{\mcitedefaultmidpunct}
{\mcitedefaultendpunct}{\mcitedefaultseppunct}\relax
\EndOfBibitem
\bibitem{Breiman}
L.~Breiman, J.~H. Friedman, R.~A. Olshen, and C.~J. Stone, {\em Classification
  and regression trees}, Wadsworth international group, Belmont, California,
  USA, 1984\relax
\mciteBstWouldAddEndPuncttrue
\mciteSetBstMidEndSepPunct{\mcitedefaultmidpunct}
{\mcitedefaultendpunct}{\mcitedefaultseppunct}\relax
\EndOfBibitem
\bibitem{AdaBoost}
R.~E. Schapire and Y.~Freund, \ifthenelse{\boolean{articletitles}}{\emph{A
  decision-theoretic generalization of on-line learning and an application to
  boosting}, }{}\href{http://dx.doi.org/10.1006/jcss.1997.1504}{Jour.\ Comp.\
  and Syst.\ Sc.\  \textbf{55} (1997) 119}\relax
\mciteBstWouldAddEndPuncttrue
\mciteSetBstMidEndSepPunct{\mcitedefaultmidpunct}
{\mcitedefaultendpunct}{\mcitedefaultseppunct}\relax
\EndOfBibitem
\bibitem{Stone:1974}
M.~Stone, \ifthenelse{\boolean{articletitles}}{\emph{Cross-validatory choice
  and assessment of statistical predictions}, }{}Journal of the Royal
  Statistical Society, Series B (Methodological) \textbf{36} (1974) 111\relax
\mciteBstWouldAddEndPuncttrue
\mciteSetBstMidEndSepPunct{\mcitedefaultmidpunct}
{\mcitedefaultendpunct}{\mcitedefaultseppunct}\relax
\EndOfBibitem
\bibitem{LHCB-PAPER-2015-025}
LHCb collaboration, R.~Aaij {\em et~al.},
  \ifthenelse{\boolean{articletitles}}{\emph{{Measurement of the decay
  $\overline{B}^0 \to D^{*+} \tau^{-}\overline{\nu}_{\tau}$}},
  }{}\href{http://arxiv.org/abs/1506.08614}{{\tt arXiv:1506.08614}}, {to appear
  in Phys. Rev. Lett.}\relax
\mciteBstWouldAddEndPunctfalse
\mciteSetBstMidEndSepPunct{\mcitedefaultmidpunct}
{}{\mcitedefaultseppunct}\relax
\EndOfBibitem
\bibitem{Wilks:1938dza}
S.~S. Wilks, \ifthenelse{\boolean{articletitles}}{\emph{{The large-sample
  distribution of the likelihood ratio for testing composite hypotheses}},
  }{}\href{http://dx.doi.org/10.1214/aoms/1177732360}{Annals Math.\ Statist.\
  \textbf{9} (1938) 60}\relax
\mciteBstWouldAddEndPuncttrue
\mciteSetBstMidEndSepPunct{\mcitedefaultmidpunct}
{\mcitedefaultendpunct}{\mcitedefaultseppunct}\relax
\EndOfBibitem
\bibitem{Skwarnicki:1986xj}
T.~Skwarnicki, {\em {A study of the radiative cascade transitions between the
  Upsilon-prime and Upsilon resonances}}, PhD thesis, Institute of Nuclear
  Physics, Krakow, 1986,
  {\href{http://inspirehep.net/record/230779/}{DESY-F31-86-02}}\relax
\mciteBstWouldAddEndPuncttrue
\mciteSetBstMidEndSepPunct{\mcitedefaultmidpunct}
{\mcitedefaultendpunct}{\mcitedefaultseppunct}\relax
\EndOfBibitem
\bibitem{LHCb-PAPER-2014-063}
LHCb collaboration, R.~Aaij {\em et~al.},
  \ifthenelse{\boolean{articletitles}}{\emph{{Study of the rare $B_s^0$ and
  $B^0$ decays into the $\pi^+\pi^-\mu^+\mu^-$ final state}},
  }{}\href{http://dx.doi.org/10.1016/j.physletb.2015.02.010}{Phys.\ Lett.\
  \textbf{B743} (2015) 46}, \href{http://arxiv.org/abs/1412.6433}{{\tt
  arXiv:1412.6433}}\relax
\mciteBstWouldAddEndPuncttrue
\mciteSetBstMidEndSepPunct{\mcitedefaultmidpunct}
{\mcitedefaultendpunct}{\mcitedefaultseppunct}\relax
\EndOfBibitem
\bibitem{Bobeth:2010wg}
C.~Bobeth, G.~Hiller, and D.~van Dyk,
  \ifthenelse{\boolean{articletitles}}{\emph{{The benefits of $\bar{B}
  \rightarrow \bar{K}^* l^+ l^-$ decays at low recoil}},
  }{}\href{http://dx.doi.org/10.1007/JHEP07(2010)098}{JHEP \textbf{1007} (2010)
  098}, \href{http://arxiv.org/abs/1006.5013}{{\tt arXiv:1006.5013}}\relax
\mciteBstWouldAddEndPuncttrue
\mciteSetBstMidEndSepPunct{\mcitedefaultmidpunct}
{\mcitedefaultendpunct}{\mcitedefaultseppunct}\relax
\EndOfBibitem
\bibitem{Bourrely:2008za}
C.~Bourrely, I.~Caprini, and L.~Lellouch,
  \ifthenelse{\boolean{articletitles}}{\emph{{Model-independent description of
  $B \to \pi \ell \nu$ decays and a determination of $|\Vub|$}},
  }{}\href{http://dx.doi.org/10.1103/PhysRevD.79.013008}{Phys.\ Rev.\
  \textbf{D79} (2009) 013008}, \href{http://arxiv.org/abs/0807.2722}{{\tt
  arXiv:0807.2722}}\relax
\mciteBstWouldAddEndPuncttrue
\mciteSetBstMidEndSepPunct{\mcitedefaultmidpunct}
{\mcitedefaultendpunct}{\mcitedefaultseppunct}\relax
\EndOfBibitem
\bibitem{Imsong:2014oqa}
I.~S. Imsong, A.~Khodjamirian, T.~Mannel, and D.~van Dyk,
  \ifthenelse{\boolean{articletitles}}{\emph{Extrapolation and unitarity bounds
  for the $b \to \pi$ form factor},
  }{}\href{http://dx.doi.org/10.1007/JHEP02(2015)126}{JHEP \textbf{02} (2015)
  126}, \href{http://arxiv.org/abs/1409.7816}{{\tt arXiv:1409.7816}}\relax
\mciteBstWouldAddEndPuncttrue
\mciteSetBstMidEndSepPunct{\mcitedefaultmidpunct}
{\mcitedefaultendpunct}{\mcitedefaultseppunct}\relax
\EndOfBibitem
\bibitem{Ball:2004ye}
P.~Ball and R.~Zwicky, \ifthenelse{\boolean{articletitles}}{\emph{{New results
  on $B \to \pi$, $K$, $\eta$ decay formfactors from light-cone sum rules}},
  }{}\href{http://dx.doi.org/10.1103/PhysRevD.71.014015}{Phys.\ Rev.\
  \textbf{D71} (2005) 014015}, \href{http://arxiv.org/abs/hep-ph/0406232}{{\tt
  arXiv:hep-ph/0406232}}\relax
\mciteBstWouldAddEndPuncttrue
\mciteSetBstMidEndSepPunct{\mcitedefaultmidpunct}
{\mcitedefaultendpunct}{\mcitedefaultseppunct}\relax
\EndOfBibitem
\bibitem{LHCb-PAPER-2014-053}
LHCb collaboration, R.~Aaij {\em et~al.},
  \ifthenelse{\boolean{articletitles}}{\emph{{Measurement of the semileptonic
  $CP$ asymmetry in $B^0$--$\overline{B}^0$ mixing}},
  }{}\href{http://dx.doi.org/10.1103/PhysRevLett.114.041601}{Phys.\ Rev.\
  Lett.\  \textbf{114} (2015) 041601},
  \href{http://arxiv.org/abs/1409.8586}{{\tt arXiv:1409.8586}}\relax
\mciteBstWouldAddEndPuncttrue
\mciteSetBstMidEndSepPunct{\mcitedefaultmidpunct}
{\mcitedefaultendpunct}{\mcitedefaultseppunct}\relax
\EndOfBibitem
\bibitem{LHCb-PAPER-2012-009}
LHCb collaboration, R.~Aaij {\em et~al.},
  \ifthenelse{\boolean{articletitles}}{\emph{{Measurement of the
  $D_s^+$--$D_s^-$ production asymmetry in 7~TeV $pp$ collisions}},
  }{}\href{http://dx.doi.org/10.1016/j.physletb.2012.06.001}{Phys.\ Lett.\
  \textbf{B713} (2012) 186}, \href{http://arxiv.org/abs/1205.0897}{{\tt
  arXiv:1205.0897}}\relax
\mciteBstWouldAddEndPuncttrue
\mciteSetBstMidEndSepPunct{\mcitedefaultmidpunct}
{\mcitedefaultendpunct}{\mcitedefaultseppunct}\relax
\EndOfBibitem
\end{mcitethebibliography}

\newpage

\centerline{\large\bf LHCb collaboration}
\begin{flushleft}
\small
R.~Aaij$^{38}$, 
B.~Adeva$^{37}$, 
M.~Adinolfi$^{46}$, 
A.~Affolder$^{52}$, 
Z.~Ajaltouni$^{5}$, 
S.~Akar$^{6}$, 
J.~Albrecht$^{9}$, 
F.~Alessio$^{38}$, 
M.~Alexander$^{51}$, 
S.~Ali$^{41}$
% ...

{\footnotesize \it
% ...
$ ^{5}$Clermont Universit\'{e}, Universit\'{e} Blaise Pascal, CNRS/IN2P3, LPC, Clermont-Ferrand, France\\
$ ^{6}$CPPM, Aix-Marseille Universit\'{e}, CNRS/IN2P3, Marseille, France\\
% ...
$ ^{9}$Fakult\"{a}t Physik, Technische Universit\"{a}t Dortmund, Dortmund, Germany\\
% ...
$ ^{37}$Universidad de Santiago de Compostela, Santiago de Compostela, Spain\\
$ ^{38}$European Organization for Nuclear Research (CERN), Geneva, Switzerland\\
% ...
$ ^{41}$Nikhef National Institute for Subatomic Physics, Amsterdam, The Netherlands\\
% ...
$ ^{46}$H.H. Wills Physics Laboratory, University of Bristol, Bristol, United Kingdom\\
% ...
$ ^{51}$School of Physics and Astronomy, University of Glasgow, Glasgow, United Kingdom\\
$ ^{52}$Oliver Lodge Laboratory, University of Liverpool, Liverpool, United Kingdom\\
% ...
}
\end{flushleft}

\end{document}